\documentclass[useAMS,referee]{biom}
\usepackage{amsmath}

\newcommand\independent{\protect\mathpalette{\protect\independenT}{\perp}}
\def\independenT#1#2{\mathrel{\rlap{$#1#2$}\mkern2mu{#1#2}}} 

\newcommand{\inp}{\stackrel{p}{\longrightarrow}}

\newcommand{\Tf}{\mathcal{T}_F}
\newcommand{\pr}{\mbox{pr}}
\newcommand{\logit}{\mbox{logit}}
\newcommand{\expit}{\mbox{expit}}
\newcommand{\calM}{\mathcal{M}}
\newcommand{\Lbar}{\bar{L}}
\newcommand{\hatalpha}{\widehat{\alpha}}
\newcommand{\hatbeta}{\widehat{\beta}}
\newcommand{\hattheta}{\widehat{\theta}}
\newcommand{\hatmu}{\widehat{\mu}}
\newcommand{\hatM}{\widehat{M}}
\newcommand{\hatp}{\widehat{p}}
\newcommand{\calK}{\mathcal{K}}
\newcommand{\calG}{\mathcal{G}}
\newcommand{\hatK}{\widehat{\calK}}
\newcommand{\calV}{\mathcal{V}}
\newcommand{\sumin}{\sum^n_{i=1}}
\newcommand{\sumjc}{\sum^{c-1}_{j=1}}

\newcommand{\pbar}{\bar{p}}
\newcommand{\hatpbar}{\widehat{\pbar}}
\newcommand{\hatpsi}{\widehat{\psi}}
\newcommand{\hatpi}{\widehat{\pi}}
\newcommand{\hatphi}{\widehat{\phi}}

\title[Proportional Odds Model with Censored Time-lagged
Outcome]{Estimation of the Odds Ratio in a Proportional Odds Model
  with Censored Time-lagged Outcome in a Randomized Clinical Trial}

\author{Anastasios A. Tsiatis$^{*}$\email{tsiatis@ncsu.edu},
Marie Davidian$^{**}$\email{davidian@ncsu.edu}, and 
Shannon T. Holloway$^{***}$\email{sthollow@ncsu.edu} \\
Department of Statistics, North Carolina State University, Raleigh, NC, USA.}

\begin{document}


\pagerange{\pageref{firstpage}--\pageref{lastpage}} 
\volume{64}
\pubyear{2008}
\artmonth{December}

\doi{10.1111/j.1541-0420.2005.00454.x}


\label{firstpage}

\begin{abstract}
  In many randomized clinical trials of therapeutics for COVID-19, the
  primary outcome is an ordinal categorical variable, and interest
  focuses on the odds ratio (active agent vs. control) under the
  assumption of a proportional odds model.  Although at the final
  analysis the outcome will be determined for all subjects, at an
  interim analysis, the status of some participants may not yet be
  determined, e.g., because ascertainment of the outcome may not be
  possible until some pre-specified follow-up time.  Accordingly, the
  outcome from these subjects can be viewed as censored.  A valid
  interim analysis can be based on data only from those subjects with
  full follow up; however, this approach is inefficient, as it does
  not exploit additional information that may be available on those
  for whom the outcome is not yet available at the time of the interim
  analysis.  Appealing to the theory of semiparametrics, we propose an
  estimator for the odds ratio in a proportional odds model with
  censored, time-lagged categorical outcome that incorporates
  additional baseline and time-dependent covariate information and
  demonstrate that it can result in considerable gains in efficiency
  relative to simpler approaches.  A byproduct of the approach is a
  covariate-adjusted estimator for the odds ratio based on the full
  data that would be available at a final analysis. \\*[0.5in]
\end{abstract}



\maketitle

\section{Introduction}\label{s:intro}

The pandemic due to the novel coronavirus SARS CoV-2 has led to an
intensive effort to identify and develop treatments for COVID-19,
coordinated through the public-private partnership Accelerating
COVID-19 Therapeutic Interventions and Vaccines (ACTIV, 2021).
Within ACTIV, candidate therapeutics are prioritized and evaluated in
clinical trials orchestrated through coordinated master protocols that
incorporate a common control arm. 

The ACTIV-3b: Therapeutics for Severely Ill Inpatients With COVID-19
(TESICO) Phase 3 master protocol (ClinicalTrials.gov NCT04843761) is
focused on evaluation of multiple agents to improve outcomes for
patients with acute respiratory failure associated with COVID-19. In a
double-blind trial within this protocol, patients hospitalized with
acute respiratory distress syndrome (ARDS) are randomized to study
agents or placebo and followed for 90 days after study entry.  The
primary outcome is based on the number of days patients are out of the
hospital without need for oxygen over the 90 day period. Because these
patients are severely ill, there is a nonnegligible probability of
death. Accordingly, the primary outcome is defined as an ordinal
categorical variable with death as the worst category (Category
6). The remaining five categories reflect recovery status of the
patient at Day 90 following randomization.  Categories 1--3 are
defined by the number of consecutive days off oxygen. For Category 4,
a patient must be out of the hospital at Day 90 but either at home on
oxygen or not at home (so continuing to receive care
elsewhere). Category 5 patients are hospitalized or in hospice care at
Day 90.  Table~\ref{tab1} summarizes the definitions of the categories
and the assumed distributions of the outcome for an investigational
agent and placebo/control. The primary treatment effect estimand is
defined in the master protocol as the odds ratio (OR) of improvement
in recovery status at Day 90 for an investigational agent versus
placebo under the assumption of a proportional odds model (Agresti,
2019).  Under the assumed distributions in Table~\ref{tab1}, the
sample size of 640 was chosen to obtain $n=602$ evaluable subjects so
as to achieve 80\% power to detect an OR of 1.5 using a two-sided test
with type 1 error 0.05, accounting for loss to follow-up or withdrawn
consent.

\begin{table}[h]%
\caption{Assumed distributions of primary outcome categories in the
  TESICO study.  Entries are expressed as percentages.}
\label{tab1}
\centering
\begin{tabular}{cp{3.5in}rr} \Hline 
\multicolumn{1}{c}{Category} & \multicolumn{1}{c}{Status at 90 days}  & 
\multicolumn{1}{c}{Investigational agent}  &
                                             \multicolumn{1}{c}{Control}  \\ \hline \\
1 & At home and off oxygen, number of days $\geq$ 77 & 17.0 & 12.0 \\
2 & At home and off oxygen, number of days 49-76 & 27.7 & 23.0 \\
3 & At home and off oxygen, number of days 1-48 & 17.2 & 17.0 \\
4 & Not hospitalized AND either at home on oxygen OR not at home & 9.1  & 10.0\\
5 & Hospitalized for medical care OR in hospice care & 4.3 & 5.0 \\
6 & Dead & 24.7 & 33.0 \\*[0.05in]
 & Total & 100.00 & 100.00 \\*[0.1in] \hline 
\end{tabular}
\end{table}

The trial is monitored by a US government-convened Data and Safety
Monitoring Board (DSMB) focused on studies of COVID-19 prevention and
therapeutics of which the first author is a member, and interim
analyses are to be conducted and reviewed by the DSMB. If all
participants are followed for the full 90 days, as would be the case
at the conclusion of the trial, the categorical outcome will be known
for all, and a standard analysis based on fitting of the proportional
odds model can be carried out to draw inference on the OR of
interest. However, at the time of an interim analysis, not all
subjects will have complete follow up:  Because of staggered entry,
some will have progressed through the entire 90 day follow-up
period, and some may have not yet entered the trial.  The status
of participants who die during the 90 days (Category 6) will be known
immediately at the time of death, but the statuses of subjects not on
study for the full 90 days at the time of the analysis will not be
known. Thus, information on Category 6 will accumulate more quickly
than that on the other outcome categories.  A naive analysis based on
this differential outcome information could lead to biased inference
on the OR; thus, an approach to taking this feature into appropriate
account is required.

A simple way to circumvent this potential for biased inference is to
carry out the standard analysis based on only the data from those
participants who have been on study for the full 90 days.  Under this
convention, information on deaths occurring for subjects not on study
for 90 days would be excluded.  Clearly, however, this approach
represents an inefficient use of the available data.  There may be
information from patients who have not died but have not been followed
for the full 90 days that could be exploited to increase precision;
e.g., if at the time of analysis a participant is at day 45 of follow
up and still in the hospital, then it would be known that his/her
status at 90 days can be only in Categories 3--6.  The standard
analysis would not take advantage of such additional information.

The primary outcome in TESICO is an example of a time-lagged response
or marked point process (Huang and Louis, 1998), characterized by the
categorical outcome, $Cat$, say, referred to as the mark, and the time
from entry into the study $T$ at which the outcome is ascertained, the
point process. In TESICO, if the outcome is Category 6, so $Cat = 6$, then $T$
is the time of death; if $Cat = 1,\ldots,5$, $T=90$ days. In this
article, we take this point of view and develop an estimator for the OR in a
proportional odds model with censored, time-lagged categorical
outcome. Although we motivate and discuss the methodology in the
context of TESICO, it is applicable in any randomized trial where
ascertainment of the categorical outcome of interest is possibly
censored at the time of analysis.

In Section~\ref{model}, we define notation and assumptions, and we
discuss inference on the OR under the proportional odds model based on
full, uncensored data from an estimating equation perspective in
Section~\ref{standard}. We then exploit the theory of semiparametrics
as in Lu and Tsiatis (2011) in Section~\ref{method} to characterize a
class of estimating equations through which consistent inference on
the OR based on censored, time-lagged categorical outcome data can be
achieved; this formulation also leads to an approach to
covariate-adjusted inference when the full data are available.  In
Section~\ref{implementation} we present a practical strategy for
implementation of the proposed methods, and we report in
Section~\ref{sim} on a suite of simulation studies demonstrating their
performance and in particular their ability to yield unbiased
estimators for the OR with enhanced efficiency relative to simpler
approaches.

\vspace*{-0.15in}

\section{Statistical framework}\label{model}

Assume that the follow-up time at which status will have been
ascertained on any participant is $\Tf$; $\Tf=90$ in TESICO.  Let $A$
denote a participant's randomized treatment assignment, taking values 0
(placebo) and 1 (investigational agent), and let $Cat$ denote
the category corresponding to the participant's status, taking values
$1, \ldots, c$, where $c$ corresponds to death; $c=6$ in TESICO.  Let
$T$ be the time at which status is ascertained, so that $T$ = time of
death if $Cat=c$, and if $Cat < c$, $T$ = time $\leq \Tf$ at which
$Cat$ is determined, which can be different for different categories
in general.  In TESICO, all non-fatal outcomes are not fully
determined until $\Tf$, so $T = \Tf$ if $Cat=1,\ldots,c-1$.  Baseline
covariate information also may be recorded on each participant prior
to initiation of assigned treatment, including demographic and
physiological characteristics, information on severity of illness, and
so on.

Letting $\logit(p) = \log\{ p/(1-p) \}$, the proportional odds model assumes that 
\begin{equation}
\logit\{ \pr(Cat \leq j \, | \, A)\} = \alpha_j + \beta A,
\hspace{0.15in} j=1,\ldots,c-1.
\label{eq:propoddsmodel}
\end{equation}
Under
(\ref{eq:propoddsmodel}), the treatment effect of interest is
expressed as the treatment-specific odds ratio $\exp(\beta)$, where
$\beta$ is the log odds ratio.  Consequently, the null hypothesis of
no effect of the investigational agent is $H_0: \beta=0$.  In TESICO,
the clinically important effect represented by OR = 1.5 corresponds to
the alternative hypothesis $H_A: \beta = \log(1.5)$.

For a trial involving $n$ participants randomized to the
investigational agent with probability $\pr(A=1) = \pi$, if all $n$
participants were to have complete follow up to time $\Tf$, as would
be the case at the time of the final analysis, the full data that would be
available for inference on $\beta$ can be summarized as independent and
identically distributed (iid)
\begin{equation}
F_i = (X_i, A_i, Cat_i, T_i), \hspace{0.15in}i=1,\ldots,n.
\label{eq:fulldata}
\end{equation}
Here, the categorical outcome is known for all $n$ subjects, and
standard methodology for fitting model (\ref{eq:propoddsmodel}) can be
used to estimate $\alpha = (\alpha_1,\ldots,\alpha_{c-1})^T$ and
$\beta$ and test $H_0$; we discuss inference based on the full data
(\ref{eq:fulldata}) in Section~\ref{standard}.  Because this inference
references only the conditional distribution of $Cat$ given $A$, the
data on $X$ and $T$ are not required or used.  As demonstrated by
Benkeser et al. (2021) in the context of COVID-19 therapeutics trials,
an alternative covariate-adjusted analysis in which $X$ is taken
into account could yield substantial precision gains; we discuss
covariate adjustment in Section~\ref{method}.

At the time of an interim analysis, some participants will not have
complete follow up to time $\Tf$.  Denote by $C$ the potential
censoring time equal to, e.g, the time from study entry to the
time of the interim analysis.  If a subject has died prior to the time
of the interim analysis, then $Cat = c$ is observed, and $T$ = time of
death $\leq C$.   Otherwise, if at the time of 
interim analysis $Cat$ is not yet determined, then $T > C$.  
Thus, let $U = \min(T,C)$ be the time on study at the time of the
analysis and $\Delta = I(T \leq C)$.  $Cat$ is
observed only if $\Delta=1$.

In addition to baseline covariates $X$, intermediate, time-dependent
information may have been collected through time $U$, while a
participant is on study.  For example, an indicator of whether or not
a patient is in the hospital at $U$ and, if out of the hospital, for
how long, will be available.  At $u$ time units after study entry,
this information can be summarized as a vector of time-dependent
covariates $L(u)$, and denote the history of such time-dependent
covariates as $\Lbar(u) = \{ L(s); \,\, 0 \leq s \leq u\}$.
As we demonstrate in the sequel, although such post-randomization
information cannot be used as part of the primary final analysis based
on the full data (\ref{eq:fulldata}), it may be advantageous at an interim
analysis.

We summarize the observed data available on $n$ participants at the
time of an interim analysis as
\vspace*{-0.2in}
\begin{equation}
O_i = \{ X_i, A_i, U_i, \Delta_i, \Delta_i Cat_i, \Lbar(U_i)\},
\hspace{0.15in} i=1,\ldots,n.
\label{eq:obsdata}
\end{equation}
The statistical challenge, which we address in Section~\ref{method},
is estimation of the log odds ratio $\beta$ in the proportional odds
model (\ref{eq:propoddsmodel}) based on the observed data
(\ref{eq:obsdata}).

\vspace*{-0.15in}

\section{Inference based on the full data }\label{standard}

In Section~\ref{method}, we appeal to the theory of semiparametrics to
develop an approach to estimation of $\beta$ using the observed data
(\ref{eq:obsdata}) via solution of an estimating equation that
exploits information in the baseline and time-dependent covariates $X$
and $\Lbar( \cdot)$ to enhance efficiency.  Following the theory, this
observed-data estimating equation is formulated based on a
suitably-chosen estimating equation using the full data
(\ref{eq:fulldata}); more precisely, using the subset of the full data 
$(A_i, Cat_i)$, $i=1,\ldots,n$.   Accordingly, we begin by considering an
estimator for $\beta$ obtained by solving an estimating equation based
on these full data.  That is, we identify a $c$-dimensional 
estimating function $\calM(F; \alpha, \beta)$ satisfying
$E_{\alpha,\beta}\{ \calM(F; \alpha, \beta) \} = 0$ for all
$\alpha,\beta$, where this notation indicates expectation taken with
respect to the distribution of $F$ evaluated at $\alpha,\beta$, such
that solution of the estimating equation
\begin{equation}
\sumin \calM(F_i; \alpha, \beta) = 0
\label{eq:fullesteqn1}
\end{equation}
in $\alpha$ and $\beta$ yields consistent and asymptotically normal 
estimators $\hatalpha_F$ and $\hatbeta_F$, say.

As is well-established (Tsiatis, 2006), generically, there is a
correspondence between consistent and asymptotically normal estimators
for a finite-dimensional parameter $\theta$ in a statistical model
based on data $F_i$, $i=1,\ldots,n$, and influence functions.  Namely,
assuming that the statistical model is correctly specified, and
letting $\theta_0$ be the true value of $\theta$ generating the data,
for a consistent and asymptotically normal estimator $\hattheta$,
there exists a function $\varphi(F)$ with $E\{ \varphi(F)\} = 0$,
referred to as the influence function for $\hattheta$, such
that
\begin{equation}
n^{1/2}(\hattheta-\theta_0) = n^{-1/2} \sumin \varphi(F_i) +
o_p(1),
\label{eq:inflfunc}
\end{equation}
where $o_p(1)$ is a term that converges in probability to zero, and
$E\{ \varphi(F) \varphi(F)^T\} < \infty$ is nonsingular.  It follows
that $n^{1/2}(\hattheta - \theta_0)$ converges in distribution to a
mean-zero normal random vector with covariance matrix
$E\{ \varphi(F) \varphi(F)^T\}$.  Estimating functions can be deduced
from influence functions, as discussed below.

Identifying $\theta = (\alpha^T,\beta)^T$ and assuming that the
proportional odds model (\ref{eq:propoddsmodel}) holds with true
values $\alpha_0$ and $\beta_0$, we now propose a full-data estimating
function $\calM(F; \alpha, \beta)$ leading to an estimating equation
as in (\ref{eq:fullesteqn1}) and obtain the associated influence
function for the resulting estimators
$\hattheta=(\hatalpha_F^T,\hatbeta_F)^T$.  Because our main focus is
estimation of $\beta$, based on these, we then deduce the associated
estimating function and influence function for the resulting estimator
$\hatbeta_F$ alone.  These quantities play a prominent role in the
developments in Section~\ref{method}.

Let $R_j = I(Cat \leq  j)$, $j=1,\ldots,c-1$.  The proportional odds model
(\ref{eq:propoddsmodel}) assumes that
$$E(R_j \, | \, A) = \expit(\alpha_j + \beta A), \hspace{0.15in} j =
1,\ldots,c-1,$$ where $\expit(u) = e^u/(1+e^u)$.  As is well known,
from Chapter 4 of Tsiatis (2006), all estimating
functions for $(\alpha^T,\beta)^T$ leading to consistent and
asymptotically normal estimators can be characterized as
$$\calG(A) \left( \begin{array}{c}
R_1 - \expit(\alpha_1 + \beta A) \\
\vdots \\
R_{c-1} - \expit(\alpha_{c-1} + \beta A) \end{array} \right),$$
where $\calG(A)$ is an arbitrary $(c \times c-1)$ matrix of functions
of $A$.  The optimal choice of $\calG(A)$ leading to the most precise
estimation asymptotically is given by  $D^T(A; \alpha,\beta)
V^{-1}(\alpha,\beta)$, where  $D(A; \alpha,\beta)$ is the $(c-1 \times
c)$ gradient matrix of the elements of $R_j - \expit(\alpha_j + \beta
A)$,  $j=1,\ldots,c-1$, with respect to $\alpha$ and $\beta$, and 
$V(\alpha,\beta)$ is the conditional covariance matrix of $R_j - \expit(\alpha_j + \beta
A)$,  $j=1,\ldots,c-1$, given $A$; this choice corresponds to the
maximum likelihood estimators given by Agresti (2019). 

We propose basing the developments in Section~\ref{method} on taking
$D(A; \alpha,\beta)$ as above and choosing $V(\alpha,\beta)$ according to the so-called
``working independence'' assumption, which leads to the full-data
estimating function
\begin{equation}
\calM(F; \alpha, \beta) = 
\left( \begin{array}{c}
R_1 - \expit(\alpha_1 + \beta A) \\
\vdots \\
R_{c-1} - \expit(\alpha_{c-1} + \beta A) \\
A \sumjc \{ R_j - \expit(\alpha_j + \beta A) \} \end{array} \right)
\hspace{0.15in} (c \times 1).
\label{eq:fullestfunc}
\end{equation}
Although this is not the efficient choice, it has been observed that
the loss of efficiency relative to the optimal choice is minimal, and,
importantly, basing the approach in Section~\ref{method} on
(\ref{eq:fullestfunc})  leads to more straightforward implementation.

Appealing to standard theory of M-estimation (Stefanski and Boos, 2002), it
is straightforward that the influence function for the
estimators solving the estimating equation (\ref{eq:fullesteqn1}) with
estimating function $\calM(F; \alpha, \beta)$ as in
(\ref{eq:fullestfunc}) is given by 
\begin{equation}
-\left[ E\left\{ \left.\frac{\partial \calM(F; \alpha,\beta)
    }{ \partial^T (\alpha^T,\beta) } \right|_{\alpha=\alpha_0, \beta=\beta_0}\right\}\right]^{-1} \calM(F;
\alpha_0,\beta_0),  
\label{eq:dMtimesM}
\end{equation}
where 
$\partial \calM(F; \alpha,\beta)/ \partial^T (\alpha^T,\beta)$ $(c \times c)$ 
is the gradient matrix comprising partial derivatives of $\calM(F;
\alpha,\beta)$ with respect to $\alpha$ and $\beta$.   

Let
\begin{equation}
p_{j0} = \expit(\alpha_j), \hspace{0.1in} p_{j1} =
\expit(\alpha_j+\beta), \hspace{0.1in} \pbar_j = \pi p_{j1}(1-p_{j1}) +
(1-\pi) p_{j0} (1-p_{j0}),
\label{eq:ps}
\end{equation}
and let $p_{j0}^0$, $p_{j1}^0$, and $\pbar_j^0$ denote these
quantities evaluated at $\alpha_0, \beta_0$.  Then it is shown in
Appendix A that the influence function
for $\hatbeta_F$ only is 
\begin{equation}
\{\calV(\alpha_0,\beta_0)\}^{-1} m(F; \alpha_0,\beta_0),
\label{eq:fullIF}
\end{equation}
 where 
\begin{equation}
\calV(\alpha,\beta) = \sumjc \frac{
  \pi(1-\pi)p_{j1}(1-p_{j1})p_{j0}(1-p_{j0}) } {\pbar_j }
\label{eq:v}
\end{equation}     
and 
\begin{equation}
 m(F; \alpha,\beta) = \sumjc \frac{ A(R_j-p_{j1})(1-\pi)p_{j0}(1-p_{j0}) 
- (1-A)(R_j-p_{j0}) \pi p_{j1}(1-p_{j1})} {\pbar_j}.
\label{eq:mfull}
\end{equation}
It can be verified that
$E_{\alpha,\beta}\{ m(F; \alpha,\beta) \} = 0$.  These results
suggest that a consistent and asymptotically normal estimator for
$\beta$ that is asymptotically equivalent to $\hatbeta_F$, with
influence function (\ref{eq:fullIF}), solves the estimating equation
$\sumin m(F_i; \alpha_0,\beta) = 0$; two estimators $\hatbeta^{(1)}$
and $\hatbeta^{(2)}$ are asymptotically equivalent if
$n^{1/2}(\hatbeta^{(1)}-\hatbeta^{(2)})$ converges in probability to
zero.  Of course, $\alpha_0$ is unknown; however, we argue in
Appendix A that a consistent and asymptotically normal
estimator for $\beta$ that is asymptotically equivalent to $\hatbeta_F$
can be obtained by substituting for $\alpha_0$ any
$n^{1/2}$-consistent estimator $\hatalpha$ and solving the estimating
equation
\begin{equation}
\sumin m(F_i; \hatalpha,\beta) = 0,
\label{eq:fullesteqnbeta}
\end{equation}
and the resulting estimator $\hatbeta$ for $\beta$ has influence function
(\ref{eq:fullIF}), from which the asymptotic variance of $\hatbeta$
can be derived.  

\section{Inference based on the observed data}\label{method}

Armed with the full data estimating function $m(F; \alpha,\beta)$ in
(\ref{eq:mfull}), we are now in a position to characterize a class of
estimating functions yielding estimators for $\beta$ based on the
observed data (\ref{eq:obsdata}).   Letting ``$\independent$'' denote
``independent of,'' we make the following two assumptions:
$$\mbox{(i)    }X \independent A, \hspace*{0.2in} 
    \mbox{(ii)   }C \independent \{X, T, Cat, \Lbar(T)\} \, | \, A,$$
where (i) is immediate because of randomization, and (ii) states that
censoring is independent of the other data conditional on $A$.  

Under (i) and (ii) and the assumed proportional odds model, with no
further assumptions, we follow Lu and Tsiatis (2011),
who invoked the theory of semiparametrics and reasoning analogous to
the above.  Given a particular full-data estimating function
$m(F; \alpha,\beta)$, which we take to be that in (\ref{eq:mfull}) here,
all consistent and asymptotically normal estimators for $\beta$ based
on the observed data (\ref{eq:obsdata}) are characterized by
estimating functions of the form
\begin{align}
m^*(O; \alpha, \beta) = \frac{\Delta m(F; \alpha,\beta)}{\calK(U,A)} -
  (A-\pi) f(X) + \int dM_c(u,A) \, h\{ u, X, A, \Lbar(u)\}.
\label{eq:mobs}
\end{align}
In (\ref{eq:mobs}), $\calK(u,a) = \pr(C \geq u\,|\,A=a)$, $u\geq0$,
$a=0,1$, is the treatment-specific survival distribution for the
censoring variable $C$; 
$$dM_c( u,a) = \{ dN_c(u) - d\Lambda_c(u,a) Y(u)\} I(A=a),$$
where $N_c(u) = I(U \leq u,\Delta=0)$ and $Y(u) = I(U \geq u)$ are the
censoring counting and at-risk processes, and
$\Lambda_c(u,a) = -\log\{ \calK(u,a)\}$ is the cumulative hazard
function for censoring; and $f(X)$ is an arbitrary function of $X$ and
$h\{ u, X, A, \Lbar(u)\}$ is an arbitrary function of $u$, $X$, $A$, and
$\Lbar(u)$.  If $\calK(u,a)$ were known, then, analogous to
(\ref{eq:fullesteqnbeta}), we could obtain an observed-data estimator
for $\beta$ as the solution to the estimating equation
\begin{equation}
\sumin m^*(O_i; \hatalpha, \beta) = 0,
\label{eq:obsesteqnbeta}
\end{equation}
where $\hatalpha$ is an $n^{1/2}$-consistent estimator for $\alpha$.
Estimators arising from solving (\ref{eq:obsesteqnbeta}) with an
estimating function like those in the class of estimating functions in
(\ref{eq:mobs}) are referred to as augmented inverse probability
complete case (AIPWCC) estimators.  If $f(\cdot)$ and
$h\{ \cdot , X, A, \Lbar(\cdot)\}$ are chosen to be equal to zero,
then the resulting estimator whose estimating function is the first
term on the right hand side of (\ref{eq:mobs}) is an inverse
probability weighted complete case (IPWCC) estimator.  The second and
third ``augmentation'' terms on the right hand side of (\ref{eq:mobs})
have mean zero under (i) and (ii) and serve to increase
efficiency relative to the IPWCC estimator by exploiting information
in $X$ and $\Lbar(u)$.  Namely, for each participant $i$, the first
augmentation term recovers information from participants randomized to
the treatment to which $i$ was not randomized through the baseline
covariates $X$, and the second term recovers information lost due to
censoring through $X$ and the time-dependent covariates $\Lbar(u)$.

In practice, $\calK(u,a)$, which appears in the leading term and
second augmentation term of (\ref{eq:mobs}), is not known.  We propose
estimating $\calK(u,a)$ by treatment-specific Kaplan-Meier estimators
$\hatK(u,a)$, for the censoring distributions, so for $a = 0, 1$ based
on the data $\{U_i, 1-\Delta_i\}$ from participants $i$ for whom
$A_i=a$.  Substituting $\hatK(u,a)$ in (\ref{eq:obsesteqnbeta}),
consider AIPWCC estimators for $\beta$ that solve estimating equations
of the form
\begin{equation}
\hspace*{-0.25in} \sumin \widetilde{m}(F_i; \alpha,\beta) = \sumin \left[ \frac{\Delta_i m(F_i; \hatalpha,\beta)}{\hatK(U_i,A_i)} -
    (A_i-\pi) f(X_i) + \int d\hatM_{c,i}(u,A_i) \, h\{ u, X_i, A_i, \Lbar_i(u)\}
  \right] = 0,
\label{eq:esteqnbeta}
\end{equation}
where
$d\hatM_{c,i}(u,a) = \{ dN_{c,i}(u) - d\widehat{\Lambda}_c(u,a)
Y_i(u)\} I(A_i=a)$, $N_{c,i}(u) = I(U_i \leq u,\Delta_i=0)$,
$Y_i(u) = I(U_i \geq u)$, and
$\widehat{\Lambda}_c(u,a) = -\log\{ \hatK(u,a)\}$.  From
(\ref{eq:inflfunc}), the asymptotic variance of an estimator follows
from its influence function.  Because of the
substitution of $\hatK(u,a)$, in constrast to the full-data case,
where the influence function (\ref{eq:fullIF}) is directly
proportional to the full-data estimating function (\ref{eq:mfull}),
the influence functions for estimators for $\beta$ solving
(\ref{eq:esteqnbeta}) involve terms in addition to
$\widetilde{m}(F; \alpha_0,\beta_0)$.  Accordingly, to facilitate
straightforward derivation of the asymptotic variance of the
estimators, using results from other
works (Zhao and Tsiatis, 1997; Bang and Tsiatis, 2000;  Lu and Tsiatis, 2011),
we instead consider
estimating equations that yield AIPWCC estimators for $\beta$ that are
asymptotically equivalent to those solving (\ref{eq:esteqnbeta}) but
are based directly on the influence functions associated with
(\ref{eq:esteqnbeta}).   These estimators are characterized by estimating
functions of the form
\begin{equation}
\label{eq:altestfunc}
\begin{aligned}
\frac{\Delta m(F; \alpha,\beta)}{\calK(U,A)} & 
+ \int \frac{ dM_c(u,A) \mu(m,u, A; \alpha, \beta)}{ \calK(u,A)} -   (A-\pi) f(X) \\
&\hspace{0.2in}+ \int dM_c(u,A) \, \big[  h\{ u, X, A, \Lbar(u)\} -\mu(h, u, A) \big],
\end{aligned}
\end{equation}
$$\mu(m,u,a; \alpha, \beta) = E\{ m(F; \alpha,\beta) \, | \, T \geq u,
A=a\}, \hspace*{0.1in}
\mu(h,u,a) = E[ h\{u, X, a, \Lbar(u)\} \, | \, T \geq u, A=a].$$

Estimating functions in (\ref{eq:altestfunc}) are distinguished by
different specifications of the functions $f(\cdot)$ and
$h\{ \cdot , X, A, \Lbar(\cdot)\}$.  Application of the theory of
semiparametrics (Tsiatis, 2006) and results from Lu and
Tsiatis (2011) show that the optimal choices of $f(\cdot)$
and $h\{ \cdot , X, A, \Lbar(\cdot)\}$ leading to the estimator for
$\beta$ with smallest asymptotic variance among all AIPWCC estimators
solving estimating equations using (\ref{eq:altestfunc}) are given by
\begin{equation}
f^{opt}(X; \alpha, \beta) = E\{ m(F; \alpha,\beta) \,|\, X, A=1\} -  E\{ m(F; \alpha,\beta) \,|\, X, A=0\},
\label{eq:fopt}
\end{equation}
\begin{equation}
h^{opt}\{u, X, A, \Lbar(u); \alpha, \beta\} = \frac{ E\{ m(F; \alpha,\beta) \,|\, T
  \geq u, X,  A, \Lbar(u) \} }{ \calK(u,A) }.
\label{eq:hopt}
\end{equation}
The conditional expectations in (\ref{eq:fopt}) and (\ref{eq:hopt})
that define the optimal choices are unknown in general.  Estimation of
$f^{opt}(X; \alpha, \beta)$ and $h^{opt}\{u, X, A, \Lbar(u); \alpha, \beta\}$ based on the observed
data involves positing models for these conditional expectations,
which is clearly challenging.  Accordingly, analogous to the approach
taken by Zhang, Tsiatis, and Davidian (2008) and Lu
and Tsiatis (2011), we consider approximating these
functions by linear combinations of basis functions.  Specifically, we
consider a linear subspace spanned by basis functions
$f_1(X),\ldots,f_M(X)$; namely, with $f_0(X) \equiv 1$, we approximate
$f^{opt}(X; \alpha, \beta)$ by
$$\sum^M_{m=0} \psi_m f_m(X).$$
Similarly, for $a=0, 1$ and specified basis functions
$h_\ell\{u,X,\Lbar(u)\}$, $\ell = 1,\ldots,L$, we approximate
$h^{opt}\{u, X, a, \Lbar(u); \alpha,\beta\}$ by
$$I(A = a) \sum^L_{\ell=1} \phi_{a,\ell} h_\ell\{u,X,\Lbar(u)\},
\hspace{0.1in} a=0, 1.$$ Although these representations are only
approximations to the complex optimal choices, experience in other
contexts (Zhang et al., 2008; Lu and Tsiatis, 2011) shows that the
resulting augmentation terms lead to considerable efficiency gains
over taking $f(\cdot)$ and $h\{ \cdot , X, A, \Lbar(\cdot)\}$ equal to
zero.

Substituting these linear representations in (\ref{eq:altestfunc}) leads
to the estimating functions
\begin{equation}
\label{eq:linestfunc}
\begin{aligned}
m(O; \alpha, \beta) &= \frac{\Delta m(F; \alpha,\beta)}{\calK(U,A)} 
+ \int \frac{ dM_c(u,A) \mu(m,u, A; \alpha, \beta)}{ \calK(u,A)} 
-  (A-\pi) \sum^M_{m=0} \psi_m f_m(X) \\
&\hspace{0.2in}+ \int dM_c(u,A) \, \sum^L_{\ell=1}
\phi_{A,\ell} \left[  h_\ell\{u,X,\Lbar(u)\} - \mu(h_\ell, u, A)\right].
\end{aligned}
\end{equation}
Deducing the optimal estimating function among those of form
(\ref{eq:linestfunc}) that leads to the corresponding estimator for
$\beta$ with smallest asymptotic variance reduces to finding the
values $\psi^{opt}_m$, $m=0,\ldots,M$, and $\phi^{opt}_{a,\ell}$, $a=0, 1$,
$\ell = 1,\ldots,L$, that minimize the variance of
$m(O; \alpha_0, \beta_0)$ and thus of the corresponding influence
function.  From the form of (\ref{eq:linestfunc}), minimizing the
variance is a linear least squares problem, discussed in Section~\ref{implementation}.

Based on these developments, we propose to estimate $\beta$ by
solving estimating equations characterized by the estimating function
$m(O; \alpha, \beta)$ in (\ref{eq:linestfunc}).  In particular,
substituting estimators for $\pi$, $\calK(U,A)$,
$\mu(m,u, A; \alpha, \beta)$, and $\mu(h_\ell, u, A)$, the proposed AIPWCC
estimator $\hatbeta$ solves in $\beta$ the estimating equation 
\begin{equation}
\label{eq:finalestequn}
\begin{aligned}
\sumin \widehat{m}(O_i; \hatalpha, \beta) &= \sumin \left( 
\frac{\Delta_i m(F_i; \hatalpha,\beta)}{\hatK(U_i,A_i)} 
+ \int \frac{ d\hatM_{c,i}(u,A_i) \hatmu(m,u, A_i; \hatalpha, \beta)}{ \hatK(u,A_i)} 
-  (A_i-\widehat{\pi}) \sum^M_{m=0} \psi_m f_m(X_i) \right.\\
&\hspace{0.2in}+ \left.\int d\hatM_{c,i}(u,A_i) \, \sum^L_{\ell=1} 
\phi_{A_i,\ell} \left[ h_\ell\{u,X_i,\Lbar_i(u)\} - \hatmu(h_\ell, u,
A_i)\right]  \right) = 0,
\end{aligned}
\end{equation}
where $\widehat{\pi} = n^{-1} \sumin A_i$, 
\begin{align*}
d\hatM_{c,i}(u,a) &= \{ dN_{c,i}(u) - d\widehat{\Lambda}_c(u,a)
Y_i(u)\} I(A_i=a)  \\
&= \left[  dN_{c,i}(u)  - \left\{ \frac{\sum^n_{k=1} dN_{c,k}(u)
    I(A_k=a) }{\sum^n_{k=1} Y_k(u) I(A_k=a) } \right\} Y_i(u) \right]
  I(A_i=a),
\end{align*}
$$\frac{\hatmu(m,u,a; \alpha,\beta)}{\hatK(u,a)} = \left\{ \sumin
  Y_i(u) I(A_i=a) \right\}^{-1} 
\sumin \left\{ \frac{\Delta_i m(F_i; \hatalpha,\beta)}{\hatK(U_i,a)}
  Y_i(u) I(A_i=a) \right\},$$
$$\hatmu(h_\ell, u,a) = \left\{ \sumin
  Y_i(u) I(A_i=a) \right\}^{-1} \sumin \big[ h_\ell\{u,X_i,\Lbar_i(u)\}
Y_i(u) I(A_i=a) \big].$$ 
From (\ref{eq:linestfunc}), $\widehat{m}(O_i; \hatalpha, \beta)$
involves substitution of these estimated quantities.  However, in
contrast to (\ref{eq:esteqnbeta}), because
$\widehat{m}(O_i; \hatalpha, \beta)$ has the form of
$m(O; \alpha, \beta)$ in (\ref{eq:linestfunc}), which when evaluated
at $(\alpha_0^T,\beta_0)^T$ is proportional to the influence function
for the ``ideal'' estimator that could be obtained by solving
estimating equations characterized by $m(O; \alpha, \beta)$ if $\pi$,
$\calK(U,A)$, $\mu(m,u, A; \alpha, \beta)$, and $\mu(h_\ell, u, A)$
were known, it follows from semiparametric theory that $\hatbeta$
shares the same influence function.  Specifically, the theory guarantees
that $\hatbeta$ is asymptotically equivalent to the ``ideal''
estimator, with asymptotic variance that can be estimated from
the form of (\ref{eq:finalestequn}); because the estimating function
is based directly on the influence function, there is no effect of
substituting estimators on the large sample properties.  

The foregoing developments subsume the case where the full data are in
fact available, as at the conclusion of the trial.  Here, $\Delta_i=1$,
$\hatK(U_i,A_i) = 1$, and $d\hatM_{c,i}(u,A_i)=0$, $i=1,\ldots,n$, and
(\ref{eq:finalestequn}) becomes
\vspace*{-0.2in}
\begin{equation}
\sumin \big\{
m(F_i; \hatalpha,\beta) -  (A_i-\widehat{\pi}) \sum^M_{m=0} \psi_m
f_m(X_i) \big\} = 0.
\label{eq:fulldatacovadjeqn}
\end{equation}
The estimating equation (\ref{eq:fulldatacovadjeqn}) is of the form of
those of Zhang et al. (2008) for covariate-adjusted inference; see
also Zhang and Zhang (2021).  Thus, the proposed methodology leads to
a covariate-adjusted estimator for the OR based on the full data that
has the potential to increase the precision of inference.  We discuss
this further in Sections~\ref{implementation}-\ref{discuss}.

\section{Practical implementation}\label{implementation}

Given a $n^{1/2}$-consistent estimator $\hatalpha$, the proposed
estimator $\hatbeta$ can be obtained by solving the
estimating equation (\ref{eq:finalestequn}) directly; e.g., by a
Newton-Raphson iterative algorithm.  This procedure must incorporate
estimation of the optimal choices of the parameters $\psi_m$, $m=0,\ldots,M$, and
$\phi_{a,\ell}$, $a=0, 1$, $\ell = 1,\ldots,L$.  As noted above,
the estimators for these parameters should minimize the
variance of $m(O; \alpha_0, \beta_0)$, which suggests estimating 
$\psi^{opt}_m$, $m=0,\ldots,M$, and $\phi^{opt}_{a,\ell}$, $a=0, 1$, $\ell = 1,\ldots,L$,
by linear regression, with ``dependent variable'' 
\begin{equation}
\widehat{\mathcal{Y}}_i(\beta) = \frac{\Delta_i m(F_i; \hatalpha,\beta)}{\hatK(U_i,A_i)} 
+ \int \frac{ d\hatM_{c,i}(u,A_i) \hatmu(m,u, A_i; \hatalpha, \beta)}{  \hatK(u,A_i)}
\label{eq:dependent}
\end{equation}
 and ``covariates'' $(A_i-\widehat{\pi}) f_m(X_i)$,
$m=0,\ldots,M$,
$I(A_i=0) \int d\hatM_{c,i}(u,0) [ h_\ell\{u,X_i,\Lbar_i(u)\} -
\hatmu(h_\ell, u, 0)]$, and
$I(A_i=1) \int d\hatM_{c,i}(u,1) [ h_\ell\{u,X_i,\Lbar_i(u)\} -
\hatmu(h_\ell, u, 1)]$, $\ell=1,\ldots,L$.  Thus, because the
``dependent variable'' depends on $\beta$, within the algorithm,
this regression would be carried out at each internal iteration with the
``dependent variable'' evaluated at the current iterate for
$\hatbeta$.  

We have found that the following alternative two-step strategy yields
comparable results but is simpler to implement.  
\begin{enumerate}[(1)]
\item Obtain initial estimates $\hatalpha$ and $\hatbeta_{init}$, say,
by solving an IPWCC version of the full-data estimating equations with
estimating function (\ref{eq:fullestfunc}); namely, solve jointly in $\alpha$
and $\beta$ the $c$ equations
\begin{align*}
\sumin &\frac{\Delta_i}{\hatK(U_i,A_i)} \{ R_{ji} - \expit(\alpha_j +
  \beta A_i) \} = 0, \hspace{0.2in} j=1,\ldots,c-1, \\
\sumin &\frac{\Delta_i}{\hatK(U_i,A_i) } A_i \sumjc \{ R_{ji} - \expit(\alpha_j + \beta A_i) \} =0.
\end{align*}

\item Obtain least squares estimators $\hatpsi_m$, $m=0,\ldots,M$, and
  $\hatphi_{a,\ell}$, $a=0, 1$, $\ell = 1,\ldots,L$, by regressing the ``dependent
  variable'' in (\ref{eq:dependent}) evaluated at $\hatbeta_{init}$, 
$\widehat{\mathcal{Y}}_i(\hatbeta_{init})$, on the ``covariates''
above, where, in $\widehat{\mathcal{Y}}_i(\hatbeta_{init})$, from 
(\ref{eq:mfull}), 
\begin{equation*}
 m(F; \hatalpha,\hatbeta_{init}) = \sumjc \frac{ A(R_j-\hatp_{j1})(1-\hatpi)\hatp_{j0}(1-\hatp_{j0}) 
- (1-A)(R_j-\hatp_{j0}) \hatpi \hatp_{j1}(1-\hatp_{j1})} {\hatpbar_j};
\end{equation*}
and $\hatp_{j0}$, $\hatp_{j1}$, and $\hatpbar_j$, $j=1,\ldots,c-1$,
are the quantities in (\ref{eq:ps}) evaluated at $\hatalpha$ and
$\hatbeta_{init}$.  Then obtain the ``predicted values'' 
$$Pred_i = (A_i-\widehat{\pi}) \sum^M_{m=0} \hatpsi_m f_m(X_i) + 
\int d\hatM_{c,i}(u,A_i) \, \sum^L_{\ell=1}
\hatphi_{A_i,\ell} \left[ h_\ell\{u,X_i,\Lbar_i(u)\} - \hatmu(h_\ell, u,
A_i)\right].$$
Finally, obtain the AIPWCC estimator $\hatbeta$ as the one-step update
\begin{equation}
\hatbeta = \hatbeta_{init} -
\{\calV(\hatalpha,\hatbeta_{init})\}^{-1}  n^{-1} \sumin Pred_i,
\label{eq:betahat}
\end{equation}
where, from (\ref{eq:v}), 
$$\calV(\hatalpha,\hatbeta_{init}) = 
\sumjc \frac{ \hatpi(1-\hatpi)\hatp_{j1}(1-\hatp_{j1})\hatp_{j0}(1-\hatp_{j0}) } {\hatpbar_j }.$$
The asymptotic variance for $\hatbeta$ can be estimated by 
\begin{equation}
\{\calV(\hatalpha,\hatbeta_{init})\}^{-2}  \sumin \{
\widehat{\mathcal{Y}}_i(\hatbeta_{init}) - Pred_i\}^2.
\label{eq:varhat}
\end{equation}
\end{enumerate}

A variation on this approach is to iterate step (2) a fixed number of
times or to ``convergence.''  We have found in simulations that
iterating does not yield significant improvement relative to carrying
out steps (1) and (2) a single time.  In Appendix B, we
sketch a heuristic justification for the one-step updated estimator
$\hatbeta$ in (\ref{eq:betahat}).  

The full-data covariate-adjusted analysis based on
(\ref{eq:fulldatacovadjeqn}) can be implemented using the
two-step strategy with $\Delta_i=1$, $\hatK(U_i,A_i) = 1$, and
$d\hatM_{c,i}(u,A_i)=0$, $i=1,\ldots,n$, substituted in steps (1) and
(2).  Here, then, in step (2),  $\widehat{\mathcal{Y}}_i(\hatbeta_{init}) = m(F; \hatalpha,\hatbeta_{init})$
is regressed only on the ``covariates'' $(A_i-\widehat{\pi})
f_m(X_i)$ to obtain  $\hatpsi_m$, $m=0,\ldots,M$, and the ``predicted
values'' are $Pred_i = (A_i-\widehat{\pi}) \sum^M_{m=0} \hatpsi_m
f_m(X_i)$, which are substituted in (\ref{eq:betahat}) to yield the
covariate-adjusted one-step updated estimator with asymptotic variance
estimated by (\ref{eq:varhat}).

\section{Simulation studies}\label{sim}

We conducted a suite of simulation experiments under scenarios
resembling the situation of the TESICO study 
introduced in Section~\ref{s:intro}.  Each simulation involved 5000
Monte Carlo replications, and in all cases for each of $n = 602$
simulated subjects, $A$ was generated as Bernoulli with
$\pr(A=1) = \pi = 0.5$, where $a=0\,(1)$ corresponds to placebo
(investigational agent).  To simulate full and observed data from
scenarios in which the proportional odds model
(\ref{eq:propoddsmodel}) holds, we generated $\Gamma$ such that the
distribution of $\Gamma$ given $A=0$ was $U(0,1)$, and the
distribution of $\Gamma$ given $A=1$ followed a proportional odds
model with
$\logit\{ \pr(\Gamma \leq u\,|\,A=1) \} = \logit\{ \pr(\Gamma \leq u
\,|\, A=0) \} + \beta$, where $\beta = \log($OR).  This was
accomplished by generating $\Upsilon \sim U(0,1)$ and taking
$\Gamma = (1-A) \Upsilon + A \Upsilon (1/\mbox{OR})/\{1 -\Upsilon +
\Upsilon (1/\mbox{OR})\}$.

In our initial experiment, Scenario 1 below, we assumed six categories with
probabilities as presented in Table~\ref{tab1} and took the true value
of the odds ratio OR = 1.5.  For a given simulated subject, we
generated the categorical outcome $Cat$ according to which interval
$\Gamma$ fell as determined by the cutpoints
$[0.00, 0.12, 0.35, 0.52, 0.62, 0.67, 1.00]$, which from
Table~\ref{tab1} are determined by the cumulative
probabilities for Categories 1--6 for placebo.  Thus, by the
construction of $\Gamma$, the probabilities of falling into each
category for a subject with $A=0$ are as in Table~\ref{tab1} for
placebo, and those for a subject with $A=1$ are as shown for the
investigational agent.  Categories 1--3 are based on the number of
consecutive days at home and off oxygen at day 90; thus, if
$\Gamma < 0.52$, we defined time in hospital as
$\mathcal{H} = 90\Gamma/0.52$, and the number of days at home and off
oxygen as $90 - \mathcal{H}$.  In TESICO, participants can leave and
then return to the hospital, so that their category status will not be
known until day 90.  For simplicity in the simulations, we did not
allow that possibility; however, the proposed methods are not
predicated on this simplification.  If $\Gamma < 0.52$, we took
$T = 90$.  If $0.52 \leq \Gamma < 0.62$ or $0.62 \leq \Gamma < 0.67$,
corresponding to Categories 4 and 5, again $T = 90$.  If
$\Gamma \geq 0.67$, corresponding to death, $T = $ time of death,
which was generated as $T_0 \sim U(a_0,b_0)$ if $A=0$ and
$T_1 \sim U(a_1,b_1)$ if $A=1$.  With these conventions,
$T = 90 I(\Gamma < 0.67) + \{ (1-A)T_0 + A T_1\} I(\Gamma \geq 0.67)$.
In all scenarios, we took $(a_0,b_0) = (0,30)$ and $(a_1,b_1) = (20,50)$.

To create a baseline covariate $X$ satisfying (i), we
generated $X \sim \mathcal{N}\{ \gamma(\Upsilon -0.5), 1\}$, so that
$X$ is independent of $A$, correlated with outcome, and does not
affect the proportional odds model.  We considered two time-dependent
covariates, so that $L(u) = \{L_1(u), L_2(u)\}$.  Defining
$\mathcal{T} = \mathcal{H} I(\Gamma<0.52) +90I(\Gamma \geq 0.52)$, 
$L_1(u) = I(\mathcal{T} < u)$, indicating whether or not
the subject was still in the hospital at time $u$, and
$L_2(u) = (90-\mathcal{T}) L_1(u)$, corresponding to the number of
days the subject was expected to be out of the hospital at day 90.
We took the censoring time $C \sim U(0, \zeta)$, where
$\zeta > 90$, and $U = \min(T,C)$, $\Delta = I(T \leq C)$.  
In all scenarios, $\gamma = 1.5$, and $\zeta = 135$, which  
led to roughly 50\% censored outcomes.  

For each Monte Carlo data set, we estimated $\beta$ and thus OR =
$\exp(\beta)$ several ways. For each subject, we generated both full
data $F = (X, A, Cat, T)$ as in (\ref{eq:fulldata}) and observed data
$O = \{X, A, U, \Delta, \Delta Cat, \Lbar(U)\}$ as in
(\ref{eq:obsdata}).  The former allows us to study the ideal full-data
analysis with no censoring that could be conducted at the end of the
study as a (unattainable at an interim analysis) benchmark and thus
evaluate the extent to which the proposed AIPWCC estimator for
$\beta$, and thus that for OR = $\exp(\beta)$, can recover information
and approach the efficiency of the ideal full-data maximum likelihood
estimator $\hatbeta_{ideal}$.  This estimator is based only on
$(A, Cat)$, so does not incorporate information in $X$ to increase
precision.  Thus, as further benchmarks, we also considered three
covariate-adjusted estimators for $\beta$: The maximum likelihood
estimator $\hatbeta_{ideal,adj}$ for $\beta$ in the conditional model
$\logit\{ \pr(Cat \leq j \, | \, X, A)\} = \alpha_j + \beta A + \gamma
X$; the estimator $\hatbeta_{LOR}$ for the average of
category-specific log-odds ratios of Benkeser et al. (2021), which
when the proportional odds assumption holds estimates the OR of
interest; and the one-step AIPWCC estimator $\hatbeta_{AIPW1,full}$
solving(\ref{eq:fulldatacovadjeqn}) discussed at the end of
Section~\ref{implementation}.  Based on the observed data, we
estimated $\beta$ five ways:
\begin{enumerate}[a]
\item Using the ``naive'' maximum likelihood estimator
  $\hatbeta_{naive}$ based on the
  data $(A_i, Cat_i)$ for $i$ with $\Delta_i=1$ only, which
  over-represents accumulating deaths;

\item Using the maximum likelihood estimator $\hatbeta_{90}$ based on data for all
  subjects who have had at least 90 days of follow up at the time of
  the analysis; with administrative censoring, we observe $C$, and
  these data are $(A_i, Cat_i)$ for $i$ with $C_i \geq 90$; 

\item Using the IPWCC estimator $\hatbeta_{init}$, denoted here as $\hatbeta_{IPW}$; 

\item Using the one-step AIPWCC estimator $\hatbeta_{AIPW1}$ that
uses only the augmentation term involving baseline covariate information $X$, so
taking $\phi_{a,\ell} = 0$, $a=0,1$, $\ell = 1,\ldots,L$, and estimating
only $\psi_m$, $m=0,\ldots,M$, to gain efficiency; 

\item Using the one-step AIPWCC estimator $\hatbeta$, which we denote
  here as $\hatbeta_{AIPW2}$, which uses both baseline and
  time-dependent covariate information to gain efficiency.
\end{enumerate}
For $\hatbeta_{ideal}$, $\hatbeta_{ideal,adj}$, $\hatbeta_{naive}$,
and $\hatbeta_{90}$, estimates and standard errors were obtained using
the \texttt{polr} function in the R package \texttt{MASS} (Venables
and Ripley, 2002).  For $\hatbeta_{LOR}$, estimates and standard
errors (back-calculated from confidence intervals) were obtained using
the R package \texttt{drord} (Benkeser, 2021).  Those for the proposed
IPWCC and AIPWCC estimators (3)-(5) were obtained using the R package
\texttt{tLagPropOdds} developed by the authors.

For Scenario 1, where data were generated according to the foregoing
scheme with OR = 1.5, results for estimation of OR = $\exp(\beta)$ are
presented in Table~\ref{tab2}.  Because of the exponentiation, not
unexpectedly, although the empirical distributions of the 5000
estimates of $\beta$ appear normal, those for estimates of
$\exp(\beta)$ exhibit slight skew due to a few large values;
accordingly, we present both Monte Carlo mean and median.  If the full
data were available, inference based on the benchmark estimator
$\exp(\hatbeta_{ideal})$ is unbiased as expected, with confidence
interval achieving the nominal 95\% level.  The adjusted estimator
$\exp(\hatbeta_{ideal,adj})$ is upwardly biased; this behavior is not
unexpected, as the proportional odds model is nonlinear, so the
conditional and marginal estimators do not coincide.  The
covariate-adjusted full-data estimators $\exp(\hatbeta_{LOR})$ and
$\exp(\hatbeta_{AIPW1,full})$ exhibit virtually identical performance
and offer a roughly 15\% efficiency gain over $\exp(\hatbeta_{ideal})$
with confidence intervals achieving the nominal level.  For the
attainable observed-data estimators, the $\exp(\hatbeta_{naive})$
yields substantially biased inference, as expected.  The estimator
$\exp(\hatbeta_{90})$ based on data from subjects with $C_i \geq 90$
only is consistent and yields valid confidence intervals; however, it
exhibits an over-two-fold efficiency loss relative to the fully
augmented estimator $\exp(\hatbeta_{AIPW2})$, which exploits
information from both baseline and time-dependent covariates.  The
IPWCC and AIPWCC estimators all yield unbiased inference and yield
valid confidence intervals; however, $\exp(\hatbeta_{IPW})$ and
$\exp\hatbeta_{AIPW1})$ show nontrivial efficiency loss relative to
the fully augmented estimator $\exp(\hatbeta_{AIPW2})$.  Relative to
the unachievable benchmark estimator $\exp(\hatbeta_{ideal})$, the
observed-data estimator $\exp(\hatbeta_{90})$ shows a three-fold
efficiency loss, which is expected under the generative censoring
distribution that yields only 1/3 of participants having full 90-day
follow up, and even greater inefficiency relative to
$\exp(\hatbeta_{LOR})$ and $\exp(\hatbeta_{AIPW1,full})$.  The
proposed augmented estimator $\exp(\hatbeta_{AIPW2})$ recovers a
substantial proportion of this efficiency loss, showing only a 35\%
loss relative to $\exp(\hatbeta_{ideal})$.  The sample size for TESICO
was chosen to yield approximately 80\% power to detect OR = 1.5 with a
two-sided test of $H_0$: OR = 1.0 at level of significance 0.05 at the
final analysis; i.e., based on $\hatbeta_{ideal}$; we discuss
empirical power results for Scenario 1 momentarily.

Scenario 2 is identical to Scenario 1, except that $H_0$ holds, so
that the true odds ratio OR = 1.0.  From Table~\ref{tab2}, relative
performance of the estimators is qualitatively similar to that under
Scenario 1.  The tests of $H_0$ achieve the nominal level except when
based on the ``naive'' estimator $\hatbeta_{naive}$, which is again
biased.  The test based on $\hatbeta_{ideal}$ achieves power very near
80\%, as planned, and, not surprisingly, the tests based on
$\hatbeta_{ideal,adj}$, $\hatbeta_{LOR} $, and
$\hatbeta_{AIPW1,full}$, which incorporate baseline covariate
information and are unbiased under $H_0$, are more powerful.  Among the
tests based on observed data, that based on $\hatbeta_{90}$, which
makes use only of data on participants with at least 90 days of follow
up at the time of the analysis achieves only 36\% power, whereas
exploiting covariate information from subjects for whom the outcome is
censored at the time of the analysis results in substantial power
gains.

Scenarios 3 and 4 are the same as Scenarios 1 and 2 except that we
generated data assuming a larger number (10) of categories.
Categories 1--7 are are based on the number of consecutive days at
home and off oxygen at day 90 and thus reflect partitioning the time
interval from 0 to 90 days more finely than in Scenarios 1--2, so that
the categorical outcome $Cat$ is determined according to which
interval $\Gamma$ falls as determined by the cutpoints
$[0.00, 0.06, 0.12, 0.22, 0.31, 0.39, 0.46,
0.52, 0.62, 0.67, 1.00]$.  Categories 8 and 9 correspond to
$0.52 \leq \Gamma < 0.62$ or $0.62 \leq \Gamma < 0.67$, and
$\Gamma \geq 0.67$, corresponds to Category 10 (death).  From
Table~\ref{tab2}, performance of the estimators is qualitatively
similar to that in Scenarios 1 and 2.   Here, with a larger number of
categories, $\exp(\hatbeta_{AIPW1,full})$ slightly outperforms 
$\exp(\hatbeta_{LOR})$, which is not unexpected, as the latter does
not exploit  the proportional odds assumption.  

\begin{table}[h]%
\centering
\caption{Simulation results, $n=602$, true proportional odds model. 
Entries are based on estimates of OR = $\exp(\beta)$ using the
  indicated estimator.  MC mean is the mean of 5000 Monte Carlo
  estimates; MC median is the median of 5000 Monte Carlo
  estimates; MC SD is the Monte Carlo standard deviation, Ave MC SE is
  the mean of Monte Carlo standard estimates, MC Cov is the Monte
  Carlo coverage of a nominal 95\% Wald-type confidence interval for
  $\exp(\beta)$, MC MSE ratio is the ratio of Monte Carlo mean
  square error for the indicated estimator relative to that for the
  AIPW2 estimator, and MC $\pr($reject $H_0)$ is the proportion of times a two-sided,
  level-0.05  Wald-type test of $H_0$: log(OR) = 0 based on the
  indicated estimator rejects $H_0$.}%
\label{tab2}
\begin{tabular}{lccccccccc} \Hline 
& $\hatbeta_{ideal}$ & $\hatbeta_{ideal,adj}$ & $\hatbeta_{LOR}$ &
                                                                  $\hatbeta_{AIPW1,full}$ &
$\hatbeta_{naive}$  & $\hatbeta_{90}$ & $\hatbeta_{IPW}$
& $\hatbeta_{AIPW1}$ & $\hatbeta_{AIPW2}$ \\ \hline \\
&\multicolumn{9}{c}{Scenario 1: 6 categories, $\beta = \log(1.5)$} \\

MC mean & 1.516  &  1.586  &  1.513  &  1.512 & 1.836   & 1.554 & 1.527 & 1.522 & 1.519\\
MC median & 1.498  &  1.569  &  1.499  &  1.497 & 1.786  & 1.487 & 1.492 & 1.490 & 1.498\\
MC SD &  0.224 &   0.240  &  0.208  &  0.208 & 0.426 &  0.419 & 0.302 & 0.289 & 0.253\\
Ave MC SE & 0.221 &   0.236 &   0.206 &   0.206 & 0.411 &  0.396 & 0.298 & 0.284 & 0.248\\
MC Cov & 0.950  &  0.937 &  0.947  &  0.948 & 0.874  & 0.948 & 0.951 & 0.949 & 0.951\\
MC MSE ratio & 0.785 &   1.012  &  0.679  &  0.677 & 4.579 &  2.774 & 1.429 & 1.306 & 1.000\\
MC $\pr($reject $H_0)$ & 0.792  &  0.861 &   0.844  &  0.845 & 0.739  & 0.357 & 0.543 & 0.580 & 0.696  \\*[0.1in]

&\multicolumn{9}{c}{Scenario 2: 6 categories, $\beta = \log(1.0)$} \\

MC mean & 1.010  &   1.008  &   1.007 &    1.007 &  1.200  &  1.031 &  1.015 &  1.012 &  1.008\\
MC median &  0.999  &   0.996 &   0.996  &   0.996 &  1.162  &  0.990 &  0.994 &  0.994 &  0.995\\
MC SD &  0.148  &   0.152 &    0.139  &   0.138 &  0.277 &    0.277 &  0.198 &  0.189 &  0.168\\
Ave MC SE & 0.147 &    0.150 &    0.137  &   0.136 &  0.267 &   0.262 &  0.195 &  0.186 &  0.163\\
MC Cov & 0.950 &    0.948 &    0.948 &    0.948 &  0.884 &   0.948 &  0.949 &  0.948 &  0.947\\
MC MSE ratio & 0.780  &   0.815 &    0.681  &   0.670 &  4.104  &  2.737 &  1.394 &  1.269 &  1.000\\
MC $\pr($reject $H_0)$ & 0.050  &   0.052 &    0.052 &    0.052 &  0.116 &   0.052 &  0.051 &  0.052 &  0.053  \\*[0.1in]

&\multicolumn{9}{c}{Scenario 3: 10 categories, $\beta = \log(1.5)$} \\

MC mean & 1.520  &  1.592  &   1.520   &  1.517  & 1.851   & 1.553  & 1.537  & 1.535  & 1.529\\
MC median & 1.502  &   1.574    & 1.508    & 1.505  & 1.802   & 1.505  & 1.509  & 1.507  & 1.506\\
MC SD &  0.219   &  0.234   &  0.210   &  0.202  & 0.415  &  0.398  & 0.318  & 0.307  & 0.252\\
Ave MC SE & 0.220   &  0.233  &   0.209  &   0.202  & 0.413  &  0.391  & 0.317  & 0.304  & 0.246\\
MC Cov &  0.953  &   0.937   &  0.952   & 0.952  & 0.872  &  0.950  & 0.951  & 0.953  & 0.948\\
MC MSE ratio & 0.751   &  0.982   &  0.689  &   0.640  & 4.592   & 2.512  & 1.589  & 1.479  & 1.000\\
MC $\pr($reject $H_0)$ & 0.805   &  0.876   &  0.849  &   0.861  & 0.760  &   0.365  & 0.513  & 0.545  & 0.719\\*[0.1in]

&\multicolumn{9}{c}{Scenario 4: 10 categories, $\beta = \log(1.0)$} \\

MC mean & 1.012 &  1.012  &  1.012  &  1.011 & 1.206 &  1.031 & 1.023 & 1.021 & 1.015\\
MC median & 1.002 &   1.002  &  1.003  &  1.002 & 1.179  & 0.997 & 1.005 & 1.001 & 1.000\\
MC SD &   0.145 &   0.149 &   0.144 &   0.135 & 0.268 &  0.265 & 0.210 & 0.203 & 0.168\\
Ave MC SE & 0.146  &  0.148 &   0.143  &  0.135 & 0.268 &  0.259 & 0.210 & 0.202 & 0.163\\
MC Cov & 0.952 &   0.951 &    0.951  &  0.951 & 0.885 &  0.952 & 0.953 & 0.948 & 0.949\\
MC MSE ratio & 0.749  &  0.782  &  0.729 &   0.647 & 4.031 &  2.500 & 1.570 & 1.469 & 1.000\\
MC $\pr($reject $H_0)$ &  0.048 &  0.049 &  0.049  &  0.049 & 0.115 &  0.048 & 0.047 & 0.052 & 0.051 \\*[0.05in] \hline
\end{tabular}
\end{table}

We carried out simulations under two additional scenarios under which
the proportional odds assumption does not hold.  We report on one of
these here; see Appendix C for
results of the second, for which the generative data mechanism is
different from those in this section.    In Scenario 5, we
adopted the same conditions as in Scenario 1, except that we generated
data from a proportional hazards model, where
$\pr(Cat \leq j) = 1- \delta_j^{\exp(\xi)}$, $j=1,\ldots,c-1$, say.
From Table~\ref{tab1}, for TESICO, $\pr(Cat \leq 3) = 0.52$ under
placebo, and, with OR = 1.5, $\pr(Cat \leq 3) = 0.619$ for the
investigational agent.  Accordingly, we chose $\xi$ so that
$\pr(Cat \leq 3)$ under the proportional hazards model is equal to
these same probabilities for placebo and investigational agent,
leading to $\exp(\xi) = 1.315$.  To generate data from this
proportional hazards model, we generated $\Gamma$ as
$\Gamma = (1-A) \Upsilon + A \{1-\Upsilon^{\exp(-\xi)}\}$, where
$\Upsilon \sim U(0,1)$.  All other aspects of the generative model
were identical to the above.  Based on $10^6$ simulated subjects, we
computed the approximate category-specific odds ratios for Categories 1--5
as (1.34, 1.42, 1.50, 1.58, 1.62); and we
estimated $\beta$ in the proportional odds model using
$\hatbeta_{ideal}$, obtaining a value of 1.48, which we use to
approximate the limit in probability of $\hatbeta_{ideal}$, the true
parameter being estimated.   The results, shown in Table~\ref{tab3},
are qualitatively similar to those in the other scenarios, reflecting
robustness to model misspecification.

\begin{table}[h]%
\centering
\caption{Simulation results, $n=602$, true proportional hazards
  model. Entries are as in Table~\ref{tab2}.}%
\label{tab3}
\begin{tabular}{lccccccccc} \Hline 
& $\hatbeta_{ideal}$ & $\hatbeta_{ideal,adj}$ & $\hatbeta_{LOR}$ &
          $\hatbeta_{AIPW1,full}$ &$\hatbeta_{naive}$  & $\hatbeta_{90}$ & $\hatbeta_{IPW}$
& $\hatbeta_{AIPW1}$ & $\hatbeta_{AIPW2}$ \\ \hline \\

&\multicolumn{9}{c}{Scenario 5: 6 categories, ``true'' $\beta = \log(1.48)$} \\

MC mean &  1.496 &   1.497  &   1.505   &  1.516  &1.898  &  1.538 & 1.531 & 1.531 & 1.528\\
MC median & 1.478  &   1.481  &   1.486  &   1.499  &1.842  &  1.491  &1.498  &1.498 & 1.499\\
MC SD &  0.219  &   0.220  &   0.223  &   0.224 & 0.438  &  0.407  &0.309  &0.310  &   0.275 \\
Ave MC SE &  0.218   &  0.219   &  0.222   &  0.223 & 0.425   & 0.391 & 0.300  &0.299  &0.265\\
MC Cov & 0.953  &   0.953  &   0.951  &   0.953  &0.831  &  0.951 & 0.948  &0.948  &0.946\\
MC MSE ratio &   0.620  &   0.626   &  0.648   &  0.663  &4.708  &  2.178  &1.261  &1.265  &1.000\\
MC $\pr($reject $H_0)$ & 0.768  &   0.769  &   0.770   &  0.788  &0.786  &0.345  &0.548  &0.546  &0.647
\\*[0.05in]  \hline 
\end{tabular}
\end{table}

\section{Discussion}\label{discuss}

We have proposed an approach to inference on the odds ratio in an
assumed proportional odds model from a randomized clinical trial with
censored time-lagged categorical outcome, as would be the case at the
time of an interim analysis.  The methods exploit baseline and
time-varying information from participants for whom the outcome of
interest is censored at the time of the analysis to achieve what can
be substantial efficiency gains over inference based on the usual
estimator for the odds ratio using only subjects with complete follow
up at the time of the analysis.  Although the methodology is motivated
in the context of COVID-19 therapeutics trials, it is applicable to
any setting with primary analysis based on the proportional odds
assumption for which the ordinal categorical o utcome of interest may
not be ascertained for all subjects at the time of an interim analysis.

The methods can be extended to more than two treatments, including the
case where the treatments are determined by a factorial arrangement of
multiple agents, analogous to the approach in Zhang et al. (2008).
The developments are multivariate generalizations of those in
Sections~\ref{standard} and \ref{method} and Appendices A and B
and are sketched in  Appendix D.

We have taken as a starting point the assumption of a proportional
odds model in accordance with the primary analyses defined in several
ACTIV COVID-19 therapeutics protocols, including TESICO. Given that
the interim analyses in these studies will be based on this assumption
as defined in these protocols, our objective is to provide more
efficient methodology under this assumption to enable more timely
decisions.  If the proportional odds assumption does not hold, then,
as is well known, the estimator for the odds ratio under this
assumption estimates some weighted combination of category-specific
odds ratios, in a manner similar to the Mantel-Haenszel method; an
analogous interpretation holds when the assumption of proportional
hazards for a time-to-event endpoint is violated.  Although such a
weighted average may not be directly interpretable, its value likely
reflects, at least approximately, the strength of the treatment effect
of interest.  In their comprehensive study demonstrating the benefits
of covariate adjustment to improve the precision of inference on
treatment effects in a range of COVID-19 therapeutics trials, Benkeser
et al. (2021) propose alternative treatment effect estimands for
ordinal categorical outcomes that are not predicated on a parametric
assumption such as that of proportional odds and that they suggest may
be more interpretable by domain scientists.  The authors focus on the
case where the full data are available, as at the time of a final
analysis, and propose semiparametric covariate-adjusted estimators
targeting these estimands; see also Zhang and Zhang (2021).  It would
be possible to derive the full-data influence functions for these
estimators and, by developments similar to those presented here,
derive corresponding observed-data influence functions and estimators
in the case of censored, time-lagged categorical outcome.

\backmatter

\vspace{-0.25in}

\section*{Acknowledgements}

The authors are grateful to Dr. Birgit Grund for helpful feedback.

\vspace*{-0.25in}

\section*{Data Availability Statement}

Data sharing is not applicable to this article, as no datasets are
generated or analysed in this paper.  The methods developed in
the paper are proposed to enable future analyses of data from ongoing
clinical trials from which the required data are not yet fully accrued.  


\vspace*{-0.25in}

\section*{Supporting Information}

An R package, \texttt{tLagPropOdds}, implementing the methodology is
available at the Comprehensive R Archive Network (CRAN) at {\tt
https://cran.r-project.org/package=tLagPropOdds}. 

\setcounter{equation}{0}
\renewcommand{\theequation}{A.\arabic{equation}}

\section*{ Appendix A: Full data influence function for estimator for $\beta$}
\label{app1}

It is straightforward to show that 
\begin{equation}
- E_{\alpha,\beta}\left\{ \frac{\partial \calM(F; \alpha,\beta)
    }{ \partial^T (\alpha^T,\beta) } \right\} = \left( \begin{array}{cc}
B_{11} & B_{12} \\
B_{12}^T & B_{22} \end{array} \right),
\label{app:one}
\end{equation}
where $B_{11} = \mbox{diag}(\pbar_1,\ldots,\pbar_{c-1})$, $B_{12} =
\pi\{ p_{11}(1-p_{11}),\ldots, p_{c-1,1}(1-p_{c-1,1}) \}^T$, and
$B_{22} = \pi \sumjc p_{j1}(1-p_{j1})$.  The influence function for
the estimator $\hatbeta_F$ for $\beta$ is the last row of the inverse of
(\ref{app:one}) evaluated at $\alpha_0, \beta_0$ post-multiplied by $\calM(F;
\alpha_0,\beta_0)$.  Using formul{\ae} for the inverse of a
partitioned matrix, the last row of  the inverse is given by 
$$-\{\calV(\alpha_0,\beta_0)\}^{-1} \left\{
  \frac{\pi p^0_{11}(1-p^0_{11})}{\pbar_1^0},\ldots,
  \frac{\pi p^0_{c-1,1}(1-p^0_{c-1,1})}{\pbar_{c-1}^0},1\right\}^T,$$
where $\calV(\alpha,\beta)$ is defined in (\ref{eq:v}).
Post-multiplying by $\calM(F; \alpha_0,\beta_0)$, it follows by
routine algebra that the influence function for $\hatbeta_F$  is given by
$\calV(\alpha_0,\beta_0)^{-1} m(F; \alpha_0,\beta_0)$ as in 
(\ref{eq:fullIF}), where $m(F; \alpha_0,\beta_0)$ is defined in
(\ref{eq:mfull}).   

We now argue heuristically that the estimator $\hatbeta$ for $\beta$
solving the estimating equation (\ref{eq:fullesteqnbeta}) with
estimating function $m(F; \hatalpha,\beta)$ and $\hatalpha$ any
$n^{1/2}$-consistent estimator for $\alpha$ is consistent and
asymptotically normal and has the influence function in
(\ref{eq:fullIF}), where by $n^{1/2}$-consistent we mean that
$n^{1/2}(\hatalpha-\alpha_0)$ is bounded in probability.  It can be
shown that
\begin{equation}
E\left\{\left. \frac{\partial m(F; \alpha,\beta) }{\partial \beta} 
  \right|_{\alpha=\alpha_0,\beta=\beta_0}\right\} =
-\calV(\alpha_0,\beta_0),
\label{app:two}
\end{equation}
\begin{equation}
\hspace*{-0.3in}E\left\{\left. \frac{\partial m(F; \alpha,\beta) }{\partial \alpha_j}
  \right|_{\alpha=\alpha_0,\beta=\beta_0}\right\} = 
-\frac{\pi(1-\pi)p^0_{j1}(1-p^0_{j1})p_{j0}^0(1-p^0_{j0})}{\pbar_j^0}+
\frac{\pi(1-\pi)p^0_{j1}(1-p^0_{j1})p_{j0}^0(1-p^0_{j0})}{\pbar_j^0} = 0 
\label{app:three}
\end{equation}
for $j=1,\ldots,c-1$.  By a standard Taylor series expansion and under regularity conditions
\begin{equation}
\label{eq:taylor}
\begin{aligned}
n^{-1/2} \sumin m(F_i; \hatalpha,\hatbeta) &\approx n^{-1/2} \sumin m(F_i; \alpha_0,\beta_0) + 
 \left\{ n^{-1} \sumin \left.\frac{\partial m(F_i; \alpha,\beta)
    }{\partial \beta}\right|_{\alpha=\alpha_0,\beta=\beta_0} \right\}
n^{1/2}(\hatbeta-\beta_0) \\
&\hspace*{0.3in}+ \left\{ n^{-1} \sumin \left.\frac{\partial m(F_i; \alpha,\beta)
    }{\partial \alpha^T}\right|_{\alpha=\alpha_0,\beta=\beta_0} \right\}
n^{1/2}(\hatalpha-\alpha_0). 
\end{aligned}
\end{equation}
The averages in braces in the second and third terms on the right hand
side converge in probability to their expectations, which for the
third term is equal to zero from (\ref{app:three}).  Substituting
these expectations, which are given in (\ref{app:two}) and
(\ref{app:three}), and rearranging, it is straightforward to show that
$\hatbeta$ has influence function (\ref{eq:fullIF}), demonstrating the
result and implying that $\hatbeta_F$ and $\hatbeta$ are
asymptotically equivalent.  This result can also be deduced directly
from semiparametric theory with $\alpha$ viewed as a nuisance
parameter using (\ref{app:two}) and (\ref{app:three}) and appealing to
the developments in Chapter 3 of Tsiatis (2006).

\setcounter{equation}{0}
\renewcommand{\theequation}{B.\arabic{equation}}

 \section*{ Appendix B: Heuristic justification for the one-step updated estimator}
\label{app2}

We first find the influence function for $\hatbeta_{init}$ solving the
estimating equations in step (1) of the two-step algorithm, so solving
in $\beta$ 
$$\sumin \frac{\Delta_i m(F_i; \hatalpha,  \beta)}{\hatK(U_i,A_i)} = 0.$$
By a Taylor series expansion, we obtain
\begin{equation}
n^{1/2} (\hatbeta_{init}-\beta_0) \approx \left\{ -n^{-1} \sumin 
\frac{\Delta_i}{\hatK(U_i,A_i)}  \left.\frac{\partial m(F_i; \hatalpha,\beta)
    }{\partial \beta}\right|_{\beta=\beta_0} \right\}^{-1} n^{-1/2}
\sumin \frac{\Delta_i m(F_i; \hatalpha,  \beta_0)}{\hatK(U_i,A_i)}.
\label{eq:appfive} 
\end{equation}
The first term on the right hand side of (\ref{eq:appfive}) can be
shown to converge in probability to
$\{\calV(\alpha_0,\beta_0)\}^{-1}$, and the second term can be written
as
\begin{align}
n^{-1/2} \sumin \frac{\Delta_i m(F_i; \hatalpha, \beta_0)}{\calK(U_i,A_i)}
+ n^{-1/2} \sumin \left\{ \frac{\Delta_i }{\hatK(U_i,A_i)} -
  \frac{\Delta_i }{\calK(U_i,A_i)} \right\} m(F_i; \hatalpha,
  \beta_0).
\label{eq:appsix}
\end{align}
Taking a Taylor series of the first term of (\ref{eq:appsix}) in $\hatalpha$ about
$\alpha_0$, using results in Zhao and Tsiatis (1997) 
to express the second term in (\ref{eq:appsix}) as 
$$n^{-1/2} \sumin \int \frac{ dM_{c,i}(u,A_i) \mu(m,u, A_i;
  \alpha_0, \beta_0)}{ \calK(u,A_i)} + o_p(1),$$
and using the fact that 
$$ n^{-1} \sumin \frac{\Delta_i }{\calK(U_i,A_i)} \left.\frac{\partial m(F_i; \hatalpha,\beta)
    }{\partial \alpha^T}\right|_{\alpha=\alpha_0,\beta=\beta_0} \inp
  E\left\{ \left.\frac{\partial m(F_i; \hatalpha,\beta)
    }{\partial \beta}\right|_{\beta=\beta_0} \right\} = 0,$$
the influence function can be shown to be
\begin{equation}
\{\calV(\alpha_0,\beta_0)\}^{-1} \left\{ \frac{\Delta m(F; \alpha_0,\beta_0)}{\calK(U,A)}  
+ \int \frac{ dM_c(u,A) \mu(m,u, A; \alpha_0, \beta_0)}{ \calK(u,A)} \right\}.
\label{eq:appseven}
\end{equation}

Now consider the one-step update in (\ref{eq:betahat}) and write it
equivalently as
\begin{equation}
n^{1/2}(\hatbeta - \beta_0) = n^{1/2}(\hatbeta_{init} - \beta_0) -
\{\calV(\hatalpha,\hatbeta_{init})\}^{-1}  n^{-1/2} \sumin Pred_i.
\label{eq:appeight}
\end{equation}
From the definition of $Pred_i$, 
\begin{align*}
n^{-1/2} \sumin Pred_i = 
n^{-1/2} &\sumin \left(  (A_i-\pi) \sum^M_{m=0} \psi^{opt}_m f_m(X_i) \right.\\
&+ \left.\int d\hatM_{c,i}(u,A_i) \, \sum^L_{\ell=1}
\phi^{opt}_{A_i,\ell} \left[ h_\ell\{u,X_i,\Lbar_i(u)\} - \hatmu(h_\ell, u,
A_i)\right] \right) + o_p(1),
\end{align*}
and $\{\calV(\hatalpha,\hatbeta_{init})\}^{-1} \inp
\{\calV(\alpha_0,\beta_0)\}^{-1}$.  Consequently, from
(\ref{eq:appeight}), the influence function of the one-step estimator
$\hatbeta$ is 
\begin{align*}
\{\calV(\alpha_0,\beta_0)\}^{-1} \left( 
\vphantom{\int dM_c(u,A) \, \sum^L_{\ell=1}
\phi^{opt}_{A,\ell} \left[  h_\ell\{u,X,\Lbar(u)\} - \mu(h_\ell, u,
  A)\right]}
\frac{\Delta m(F; \alpha_0,\beta_0)}{\calK(U,A)} \right.
&+ \int \frac{ dM_c(u,A) \mu(m,u, A; \alpha_0, \beta_0)}{ \calK(u,A)} 
-  (A-\pi) \sum^M_{m=0} \psi^{opt}_m f_m(X)\\
&\hspace{0.2in}- \left. \int dM_c(u,A) \, \sum^L_{\ell=1}
\phi^{opt}_{A,\ell} \left[  h_\ell\{u,X,\Lbar(u)\} - \mu(h_\ell, u,
  A)\right] \right),
\end{align*}
which is the influence function of estimators solving the optimal
estimating equation of form (\ref{eq:finalestequn}).

\setcounter{equation}{0}
\renewcommand{\theequation}{C.\arabic{equation}}

\section*{ Appendix C: Additional simulation results}
\label{app3}

Here, we report on an additional simulation scenario under which the
proportional odds assumption does not hold, which we refer to as
Scenario 6.  We generated data according to a different mechanism that
creates the outcome prospectively rather than through a latent
variable ($\Upsilon$) as in the scenarios discussed in the main paper,
again with six categories, with $Cat=6$ corresponding to death within
90 days.  As in the main paper, we simulated 5000 Monte Carlo data
sets, each involving $n = 602$ simulated subjects.

We took the hazard rate for leaving the hospital for placebo
to be constant and equal to $\lambda_{H0} = 0.0149$ for the first 90
days and the hazard rate for death for placebo within the first 40
days to be $\lambda_{D0} = 0.0139$; these values were chosen to
correspond to the probability of leaving the hospital by 90 days
without dying to be 0.52 and the probability of dying prior to leaving
the hospital within 40 days to be 0.33.  To characterize Categories 1
to 3, we defined cutpoints $c_1 = 9.16$ and $c_2 = 39.19$ so that the
probabilities of leaving the hospital without dying by $c_1$ or $c_2$
were 0.12 and 0.35, respectively.  Under this scenario, about 15\% of
placebo subjects are still hospitalized and alive at 90 days.  For
placebo subjects remaining in the hospital at 90 days, we assigned
them to Categories 4 and 5 with probabilities $p_{40} = 2/3$ and
$1-p_{40}$, so that 10\% and 5\% were in Categories 4 and 5,
respectively.  We generated subjects under treatment such that the
hazard for leaving the hospital was
$\lambda_{H1} = \lambda_{H0} \exp(\xi)$ for the first 90 days,
$\xi = \log(1/0.775)$, so that the hazard increased relative to
placebo.  Likewise, we took the hazard of death within the first 90
days to be $\lambda_{D1} = \lambda_{D0} \exp(-\xi)$, so that the
hazard decreased relative to placebo.  Finally, we took the
probabilities of assignment to Categories 4 and 5 to be $p_{41}$ and
$(1-p_{41})$, where $\logit(p_{41}) = \logit(p_{40}) + \xi$, increasing
the probability of being in Category 4.  

To generate a subject, we first generated $A$ as Bernoulli with
$\pi = 0.5$ and generated $U_{H}$,
$U_{D}$, and $U_{4}$ to be independent $U(0,1)$.  We then generated
for each of $a=0, 1$ $\chi_{Ha} = -log(U_{H})/\lambda_{Ha}$, so as exponential
with hazard rate $\lambda_{Ha}$; and
$\chi_{Da}=-log(U_{D})/\lambda_{Da}$, so as exponential with hazard
rate $\lambda_{Da}$.  Note that $\chi_{Ha}$, $a = 0, 1$, are
correlated; and similarly for $\chi_{Da}$.  Then, if
$\chi_{Da} < \min(\chi_{Ha},40)$, we set $Cat_a = 6$, and
$T_a = \chi_{Da}$. Otherwise, if $\chi_{Da} \leq \min(\chi_{Ha},40)$,
we set $Cat_a = 1$ if $\chi_{Ha} < 9.16$, $Cat_a = 2$ if
$9.16 \leq \chi_{Ha} < 39.19$, or $Cat_a = 3$ if
$39.19 \leq \chi_{Ha} < 90$, and $T_a = \chi_{Ha}$.  If $\chi_{Ha} > 90$, we generated
$B_{4a} = I(U_{4} \leq p_{4a})$, so that $B_{4a}$ are correlated, and
set $Cat_a = 4$ or $Cat_a=5$ as $B_a = 1$ or 0, respectively, and $T_a=90$.  We then
obtained $Cat = A Cat_1 + (1-A)Cat_0$ and
$T = 90 I(Cat < 6) + \{A T_1 + (1-A)T_0\} I(Cat = 6)$.  Based on
$10^6$ simulated subjects, this 
generative model results in approximate category-specific odds ratios
of $(1.33, 1.48, 1.60, 1.56,1.47)$ corresponding to categories
$(1,\ldots,5)$.  

To create a baseline covariate $X$ satisfying assumption (i) in
Section 4 of the main paper, we took
$X\sim \mathcal{N}\{ \gamma( Cat_0-3.5), 1\}$, so that $X$ is
independent of $A$ and correlated with outcome by virtue of the
mechanism above, , with $\gamma = 0.25$.  We took the censoring time
$C \sim U(0, \zeta)$, where $\zeta = 135$, and $U = \min(T,C)$,
$\Delta = I(T \leq C)$, corresponding to about 50\% censoring.
Finally, we created two time-dependent covariates, so that
$L(u) = \{L_1(u), L_2(u)\}$, by defining
$\mathcal{T} = \{A T_1 + (1-A)T_0\} I(Cat<=3) + 90 I(Cat>3)$ and
setting $L_1(u) = I(\mathcal{T} < u)$, indicating whether or not
the subject was still in the hospital at time $u$, and
$L_2(u) = (90-\mathcal{T}) L_1(u)$, corresponding to the number of
days the subject was expected to be out of the hospital at day 90.

Because the proportional odds assumption does not hold, as in Scenario
5 in the main paper, we approximated the limit in probability of
$\hatbeta_{ideal}$, the true parameter being estimated, by simulating
$10^6$ subjects and obtaining the Monte Carlo average of estimates
$\hatbeta_{ideal}$, 1.49. 

Table~\ref{tab6} presents the results.  As for Scenario 5 in the main
paper under a true proportional hazards assumption, the results
reflect robustness to model misspecification. 

\begin{table}[h]%
\centering
\caption{Simulation results, $n=602$, Scenario 6.  Entries are as in
  Table~\ref{tab2} of the main paper.}%
\label{tab6}
\begin{tabular}{lccccccccc} \Hline 
& $\hatbeta_{ideal}$ & $\hatbeta_{ideal,adj}$ & $\hatbeta_{LOR}$ &
          $\hatbeta_{AIPW1,full}$ &$\hatbeta_{naive}$  & $\hatbeta_{90}$ & $\hatbeta_{IPW}$
& $\hatbeta_{AIPW1}$ & $\hatbeta_{AIPW2}$ \\ \hline \\

&\multicolumn{9}{c}{Scenario 6: 6 categories, ``true'' $\beta = \log(1.49)$} \\

MC mean &  1.504 &  1.569   &   1.501   &   1.519  & 1.556   &  1.539  & 1.535 & 1.533  & 1.528\\
MC median & 1.487   &   1.550  &    1.487   &   1.503  & 1.518   &  1.489  & 1.502  & 1.507  & 1.506\\
MC SD &  0.222   &   0.238   &   0.207  &    0.210  & 0.352  &   0.407  & 0.298  & 0.285  & 0.245\\
Ave MC SE &  0.220  &    0.234  &    0.204  &    0.207  & 0.344  &   0.393  & 0.291  & 0.277  & 0.238\\
MC Cov &  0.949   &   0.941  &    0.952  &    0.949  & 0.947  &   0.951  & 0.945  & 0.946  & 0.943\\
MC MSE ratio & 0.805  &    1.023    &  0.700  &    0.730  & 2.087 &    2.739  & 1.476  & 1.351 &  1.000\\
MC $\pr($reject $H_0)$ & 0.774  &    0.839  &    0.834  &    0.858  & 0.474  &   0.347 &  0.571  & 0.616  & 0.749\\*[0.05in]  \hline 
\end{tabular}
\end{table}

\setcounter{equation}{0}
\renewcommand{\theequation}{D.\arabic{equation}}

\section*{ Appendix D: Extension to more than two treatments}
\label{app4}

We consider the case of $K \geq 2$ treatments.  The full data $F_i$ in 
(\ref{eq:fulldata}) and observed data $O_i$ in (\ref{eq:obsdata}) are the same
except that $A$ takes on values $a=0, \ldots, K-1$, where 0 indicates
placebo and $1, \ldots, K-1$ indicate investigational agents.  Subjects are
randomized to the $K$ treatments with probabilities $\pr(A=a) =
\pi_a$, where $\sum_{a=0}^{K-1} \pi_a = 1$.  The proportional odds
model is now
\begin{equation}
\logit\{ \pr(Cat \leq j \, | \, A)\} = \alpha_j + \beta_1 I(A=1) +
\cdots + \beta_{K-1} I(A=K-1), 
\hspace{0.15in} j=1,\ldots,c-1,
\label{mult:propoddsmodel}
\end{equation}
where, for the $a$th investigational agent, $a=1, \ldots, K-1$, the
treatment effect is expressed as the treatment-specific odds ratio
$\exp(\beta_a)$; and $\alpha_j$, $j=1,\ldots, c-1$, are as in
(\ref{eq:propoddsmodel}).  Defining now $\beta =
(\beta_1,\ldots,\beta_{K-1})^T$, identify $\theta$ in (\ref{eq:inflfunc})
as $\theta = (\alpha^T,\beta^T)^T$, with true values
$\theta_0=(\alpha_0^T,\beta_0^T)^T$.  

Analogous to (\ref{eq:fullestfunc}), consider the full-data
estimating function for $\theta$ based on the ``working
independence'' assumption given by 
\begin{equation}
\calM(F; \alpha, \beta) = 
\left( \begin{array}{c}
R_1 - \expit\{\alpha_1 + \beta_1  I(A=1) +
\cdots + \beta_{K-1} I(A=K-1)\}\\
\vdots \\
R_{c-1} - \expit\{\alpha_{c-1} + \beta_1 I(A=1) +
\cdots + \beta_{K-1} I(A=K-1) \} \\
I(A=1) \sumjc \{ R_j - \expit(\alpha_j + \beta_1) \} \\
\vdots \\
I(A=K-1) \sumjc \{ R_j - \expit(\alpha_j + \beta_{K-1}) \} 
\end{array} \right)
\hspace{0.15in} (c-1 + K-1 \times 1).
\label{mult:fullest}
\end{equation}
As before, denote by $\hattheta^F=(\hatalpha^{F\,T},
\hatbeta^{F\,T})^T$ the estimators obtained by solving (\ref{eq:fullesteqn1}) with $\calM(F;
\alpha, \beta)$ as in (\ref{mult:fullest}).    Analogous to
(\ref{eq:dMtimesM}), the influence function
for $\hattheta^F$ is then 
\begin{equation}
-\left[ E\left\{ \left.\frac{\partial \calM(F; \alpha,\beta)
    }{ \partial^T (\alpha^T,\beta^T) } \right|_{\alpha=\alpha_0, \beta=\beta_0}\right\}\right]^{-1} \calM(F;
\alpha_0,\beta_0), 
\label{mult:dMtimesM}
\end{equation}
and, analogous to (\ref{app:one}), the gradient matrix in
(\ref{mult:dMtimesM}) satisfies
$$- E_{\alpha,\beta}\left\{ \frac{\partial \calM(F; \alpha,\beta)
    }{ \partial^T (\alpha^T,\beta^T) } \right\} = \left( \begin{array}{cc}
B_{11} & B_{12} \\
B_{12}^T & B_{22} \end{array} \right), \hspace*{0.1in} (c-1+K-1
\times c-1+K-1),$$
where now, as in (\ref{eq:ps}), $p_{j0} = \expit(\alpha_j)$, $p_{ja} =
\expit(\alpha_j+\beta_a)$, and 
$\pbar_j = \sum^{K-1}_{a=0} \pi_a p_{ja}(1-p_{ja})$, $j=1,\ldots,c-1$,
$a=1,\ldots,K-1$; and 
$B_{11} = \mbox{diag}(\pbar_1,\ldots,\pbar_{c-1})$ as before,  
$B_{12}$ is the $(c-1 \times K-1)$ matrix with $(j,a)$ element given
by $\pi_a p_{ja}(1-p_{ja})$, and $B_{22} = \mbox{diag}\{ \pi_1 \sum_{j=1}^{c-1}
p_{j1}(1-p_{j1}), \ldots, \pi_{K-1} \sum_{j=1}^{c-1}
p_{j,K-1}(1-p_{j,K-1}) \}$ $(K-1 \times K-1)$.  As in
Section~\ref{standard}, add a superscript 0 to these quantities to
denote evaluation at $\alpha_0,\beta_0$.  The $(K-1 \times 1)$ influence function
for $\hatbeta^F$ is then given by the last $K-1$ rows of
(\ref{mult:dMtimesM}) with these definitions, which, using formul{\ae}
for the inverse of a partitioned matrix, can be written as
$-\{\calV(\alpha_0,\beta_0)\}^{-1} m(F; \alpha_0, \beta_0)$,
where 
\begin{equation} 
\calV(\alpha,\beta) = B_{22} - B^T_{12} B_{11}^{-1} B_{12},
\hspace*{0.1in}
m(F; \alpha, \beta) = \big( B^T_{12}B_{11}^{-1} , \mathcal{I}_{K-1} \big) \calM(F; \alpha,\beta) ,
\label{mult:Va0b0}
\end{equation}
and $\mathcal{I}_{K-1}$ denotes a $(K-1 \times K-1)$ identity matrix.  

The analogs for general $K$ of the observed-data estimating functions in
(\ref{eq:mobs}) are of the form
\begin{equation}
m^*(O; \alpha, \beta) = \frac{\Delta m(F; \alpha,\beta)}{\calK(U,A)} -
\sum^{K-1}_{a=0} \{ I(A=a)-\pi_a\} f^{(a)}(X) + \int dM_c(u,A) \, h\{u, X, A, \Lbar(u)\},
\label{mult:mstar}
\end{equation}
where now $m^*(O; \alpha, \beta)$ is $(K-1 \times 1)$, and 
$$f^{(a)}(X) = \left( \begin{array}{c} f^{(a)}_1(X) \\ \vdots \\ f^{(a)}_{K-1}(X) \end{array}\right),
                    \hspace{0.1in} a=0,\ldots, K-1, \hspace{0.15in} 
h\{ u, X, A, \Lbar(u)\} = \left( \begin{array}{c} h_1\{ u, X,A,\Lbar(u)\} \\ \vdots 
\\ h_{K-1}\{ u, X, A, \Lbar(u)\} \end{array} \right).$$
The optimal choices of $f^{(a)}( \cdot )$ and $h\{ \cdot , X, A,
\Lbar(\cdot)\}$ are, analogous to (\ref{eq:fopt}) and (\ref{eq:hopt}),
$$f^{(a)opt}(X; \alpha, \beta) = E\{ m(F; \alpha,\beta) \,|\, X,A=a\},
\hspace{0.1in} a=0,\ldots,K-1,$$
$$h^{opt}\{u, X, A, \Lbar(u); \alpha, \beta\} = \frac{ E\{ m(F; \alpha,\beta) \,|\, T
  \geq u, X,  A, \Lbar(u) \} }{ \calK(u,A) }.$$
With $K=2$, so that $f^{(a)}(X)$, $a=0, 1$ are scalar
quantities and with $\pi_1 = \pi$ and $\pi_0 = 1-\pi$, note that in (\ref{mult:mstar}),
$\sum^{1}_{a=0} \{ I(A=a)-\pi_a\} f^{(a)}(X) = 
(A-\pi) \{  f^{(1)}(X)- f^{(0)}(X)\} = (A-\pi) f(X)$, say, as in
(\ref{eq:mobs}).  

As for $K=2$, because modeling the conditional expectations above is
challenging, we approximate them by linear combinations of basis
functions.  Here, for each $a = 0, \ldots, K-1$, we consider the
linear subspace spanned by basis functions $f_1(X), \ldots, f_M(X)$
and, with $f_0(X) \equiv 1$, approximate the $r$th element of
$f^{(a)opt}(X)$, $r=1,\ldots,K-1$, by 
$$\sum_{m=0}^M \psi_{ar,m} f_m(X), \hspace*{0.15in} a=0,\ldots,K-1.$$
Because $\sum^{K-1}_{a=0} \{I(A=a)-\pi_a\} = 0$, and because we are
taking linear combinations of $f_0(X), \ldots, f_M(X)$, generalizing
the above when $K=2$, it suffices to consider 
$$\sum^{K-1}_{a=1}  \{I(A=a)-\pi_a\} \sum^M_{m=0} \psi_{ar,m} f_m(X)$$
for the $r$th element of the first augmentation term in
(\ref{mult:mstar}); i.e., we consider linear combinations of
$f_0(X), \ldots, f_M(X)$ only for $a = 1,\ldots,K-1$.  Similarly, for
$a=0,\ldots,K-1$ and specified basis functions 
$h_\ell\{u,X,\Lbar(u)\}$, $\ell = 1,\ldots,L$, we approximate the
$r$th element of $h^{opt}\{u, X, a, \Lbar(u); \alpha,\beta\}$ by
$$I(A = a) \sum^L_{\ell=1} \phi_{ar,\ell} h_\ell\{u,X,\Lbar(u)\},
\hspace{0.1in} r=1,\ldots,K-1.$$  Therefore, analogous to
(\ref{eq:finalestequn}), the proposed AIPWCC estimator $\hatbeta$
solves in $\beta$ the estimating equation 
$$\sumin \widehat{m}(O_i; \hatalpha, \beta) = 
\sumin \left( \begin{array}{c}  
\widehat{m}_1(O_i; \hatalpha, \beta) \\ \vdots \\
\widehat{m}_{K-1}(O_i; \hatalpha, \beta) \end{array} \right),$$
where for $r=1,\ldots,K-1$ 
\begin{equation*}
\begin{aligned}
\sumin \widehat{m}_r(O_i; \hatalpha, \beta) &= \sumin \left( 
\frac{\Delta_i m_r(F_i; \hatalpha,\beta)}{\hatK(U_i,A_i)} 
+ \int \frac{ d\hatM_{c,i}(u,A_i) \hatmu_r(m,u, A_i; \hatalpha, \beta)}{ \hatK(u,A_i)}  \right.\\
&\hspace{0.2in}-  \sum^{K-1}_{a=1} \{ I(A_i =a) - \widehat{\pi}_a\} \sum^M_{m=0} \psi_{ar,m} f_m(X_i)\\
&\hspace{0.2in}+ \left.\int d\hatM_{c,i}(u,A_i) \, \sum^L_{\ell=1} 
\phi_{A_i r,\ell} \left[ h_\ell\{u,X_i,\Lbar_i(u)\} - \hatmu(h_\ell, u,
A_i)\right]  \right),
\end{aligned}
\end{equation*}
$\widehat{\pi}_a = n^{-1} \sumin I(A_i=a)$, $a=1,\ldots,K-1$,
and 
$$\frac{\hatmu_r(m,u,a; \alpha,\beta)}{\hatK(u,a)} = \left\{ \sumin
  Y_i(u) I(A_i=a) \right\}^{-1} 
\sumin \left\{ \frac{\Delta_i m_r(F_i; \hatalpha,\beta)}{\hatK(U_i,a)}
  Y_i(u) I(A_i=a) \right\};$$
and all other quantities are as in Section~\ref{method}.  

Analogous to the two-treatment case, estimators for the optimal
choices $\psi^{opt}_{ar,m}$, $a=1,\ldots,K-1$, $m=0,\ldots,M$, and
$\phi^{opt}_{ar,\ell}$, $a=0,\ldots, K-1$, $\ell = 1,\ldots,L$,  can be
obtained by least squares regression, element-by-element for
$r=1,\ldots,K-1$.  Namely, for each $r=1,\ldots,K-1$ separately, regress the ``dependent variable'' 
$$\widehat{\mathcal{Y}}_{ir}(\beta) = \frac{\Delta_i m_r(F_i; \hatalpha,\beta)}{\hatK(U_i,A_i)} 
+ \int \frac{ d\hatM_{c,i}(u,A_i) \hatmu_r(m,u, A_i; \hatalpha, \beta)}{  \hatK(u,A_i)}$$
on ``covariates'' $\{I(A_i=a)-\widehat{\pi}_a\} f_m(X_i)$,
$m=0,\ldots,M$, $a=1,\ldots,K-1$, and 
$I(A_i=a) \int d\hatM_{c,i}(u,a) [ h_\ell\{u,X_i,\Lbar_i(u)\} -
\hatmu(h_\ell, u, a)]$, $\ell=1,\ldots,L$, $a=0,\ldots,K-1$.  

The foregoing developments lead to the following generalization of the
two-step strategy in Section~\ref{implementation}:
\begin{enumerate}[(1)]
\item Obtain initial estimates $\hatalpha$ and $\hatbeta_{init}$, say,
by solving an IPWCC version of the full-data estimating equations with
estimating function (\ref{mult:fullest}); namely, solve jointly in $\alpha$
and $\beta$ the $(c-1+K-1)$ equations
\begin{align*}
\sumin &\frac{\Delta_i}{\hatK(U_i,A_i)} \Big[ R_{ji} - \expit\{\alpha_1 + \beta_1  I(A_i=1) +
\cdots + \beta_{K-1} I(A_i=K-1)\} \Big]= 0, \\
&\hspace{2in} j=1,\ldots,c-1, \\
\sumin &\frac{\Delta_i}{\hatK(U_i,A_i) } I(A_i=a) \sumjc \Big [ R_{ji} - \expit\{\alpha_1 + \beta_1  I(A_i=1) +
\cdots + \beta_{K-1} I(A_i=K-1)\} \Big]= 0, \\ 
&\hspace{2in} a=1,\ldots,K-1.
\end{align*}

\item Obtain least squares estimators $\hatpsi_{ar,m}$, $a=1,\ldots,K-1$, 
  $m=0,\ldots,M$, and $\hatphi_{ar,\ell}$, $a=0,\ldots,K-1$, 
  $\ell = 1,\ldots,L$, for each $r = 1\ldots,K-1$ by
  regressing $\widehat{\mathcal{Y}}_{ir}(\hatbeta_{init})$ on the
  ``covariates'' above, separately for each $r$.  Then obtain the
  ``predicted values'' $Pred_i = (Pred_{i1},\ldots,Pred_{i,K-1})^T$,
  where
\begin{align*}
Pred_{ir} = \sum^{K-1}_{a=1} &\big\{ I(A_i=a) - \widehat{\pi}_a\big\}
\sum^M_{m=0} \hatpsi_{ar,m} f_m(X_i) \\
&+  \int d\hatM_{c,i}(u,A_i) \, \sum^L_{\ell=1}
\hatphi_{A_i r,\ell} \left[ h_\ell\{u,X_i,\Lbar_i(u)\} - \hatmu(h_\ell, u,
A_i)\right];
\end{align*}
and obtain the AIPWCC estimator $\hatbeta$ as the one-step update
$$\hatbeta = \hatbeta_{init} -
\{\calV(\hatalpha,\hatbeta_{init})\}^{-1}  n^{-1} \sumin Pred_i,$$
where $\calV(\alpha,\beta)$ is defined in (\ref{mult:Va0b0}).  
The asymptotic variance for $\hatbeta$ can be estimated by 
$$\{\calV(\hatalpha,\hatbeta_{init})\}^{-1}  \sumin \{
\widehat{\mathcal{Y}}_i(\hatbeta_{init}) - Pred_i\}\{
\widehat{\mathcal{Y}}_i(\hatbeta_{init}) - Pred_i\}^T \{\calV(\hatalpha,\hatbeta_{init})\}^{-1},$$
where $\widehat{\mathcal{Y}}_{i}(\hatbeta_{init}) = \{
\widehat{\mathcal{Y}}_{i1}(\hatbeta_{init}),\ldots, \widehat{\mathcal{Y}}_{i,K-1}(\hatbeta_{init})\}^T.$
\end{enumerate}

Table~\ref{multtab} shows results of simulation experiments with $K=3$
analogous to those under Scenarios 1 and 2 in Section~\ref{sim}, each
involving 5000 Monte Carlo replications with $n=903$ subjects and
$\pi_0=\pi_1=\pi_2 = 1/3$, with $A$ generated as multinomial with
these probabilities.  We generated $\Gamma$ such that the distribution
of $\Gamma$ given $A=0$ was $U(0,1)$ and those of $\Gamma$ given
$A=a$, $a = 1, 2$ followed a proportional odds model as in
(\ref{mult:propoddsmodel}) with
$\logit\{ \pr(\Gamma \geq u | A=a) \} = \logit\{ \pr(\Gamma \geq u |
A=0) \} + \beta_a$, where $\beta_a = \log(\mbox{OR}_a)$, $a = 1, 2$,
and OR$_1$ = 1.5 and OR$_2$ = 1.2 under Scenario 1 and OR$_1$ =
OR$_2$ = 1.0 under Scenario 2.  This was accomplished by taking
$\Gamma = I(A=0) \Upsilon + I(A=1) \Upsilon (1/\mbox{OR}_1)/\{1
-\Upsilon + \Upsilon (1/\mbox{OR}_1)\} + I(A=2) \Upsilon
(1/\mbox{OR}_2)/\{1 -\Upsilon + \Upsilon (1/\mbox{OR}_2)\}$.  We assumed
the six categories in Table~\ref{tab1} of the main paper and generated $\mathcal{H}$ and
$T$ as described in Section~\ref{sim}, except that when
$\Gamma \geq 0.67$, corresponding to death, we generated $T = $ time
of death by generating $T_a \sim U(a_a,b_a)$, $a = 0, 1, 2$,
$(a_0,b_0) = (0,30)$, $(a_1,b_1) = (20,50)$, $(a_2,b_2) = (25,60)$ and
taking
$T = 90 I(\Gamma < 0.67) + \{ I(A=0)T_0 + I(A=1) T_1 + I(A=2) T_2\}
I(\Gamma \geq 0.67)$.  The time-independent and time-dependent
covariates $X$ and $L_1(u)$, $L_2(u)$ and the censoring time $C$ were
generated as in Section~\ref{sim}.  For brevity, Table~\ref{multtab} shows results
for estimation of $\beta_1$ and $\beta_2$ using methods c-e in
Section~\ref{sim}; i.e., using the IPWCC estimator $\hatbeta_{IPW}$  and the one-step
AIPWCC estimators $\hatbeta_{AIPW1}$ and $\hatbeta_{AIPW2}$ using only
the augmentation term involving baseline covariates (so with
$\phi_{ar,\ell} = 0$, $a=0, 1, 2$, $\ell= 1,\ldots,L$, $r =
1,\ldots,K-1$)  and using both augmentation terms.   Overall, the results are 
qualitatively similar to those in Section~\ref{sim} and demonstrate the
good performance of the methods and the considerable efficiency gains
possible via the fully augmented estimator $\hatbeta_{AIPW2}$.    

\begin{table}[h]%
\centering
\caption{Simulation results, $n=903$, true proportional odds model
  with $K=3$.  Entries are based on estimates of OR$_1$ = $\exp(\beta_1)$ and
  OR$_2$ = $\exp(\beta_2)$ using the indicated estimator.  MC mean is
  the mean of 5000 Monte Carlo estimates; MC median is the median of
  5000 Monte Carlo estimates; MC SD is the Monte Carlo standard
  deviation, Ave MC SE is the mean of Monte Carlo standard estimates,
  MC Cov is the Monte Carlo coverage of a nominal 95\% Wald-type
  confidence interval for $\exp(\beta_a)$, $a = 1, 2$, MC MSE ratio is the ratio of
  Monte Carlo mean square error for the indicated estimator relative
  to that for the AIPW2 estimator, and MC $\pr($reject $H_0)$ is the
  proportion of times a two-sided, level-0.05 Wald-type test of $H_0$:
  log(OR) = 0 based on the indicated estimator rejects $H_0$.}
\label{multtab}
\begin{tabular}{lcccccc} \Hline 
& $\hatbeta_{1,IPW}$ & $\hatbeta_{1,AIPW1}$ & $\hatbeta_{1,AIPW2}$ &
                                                                     $\hatbeta_{2,IPW}$
  & $\hatbeta_{2,AIPW1}$ & $\hatbeta_{2,AIPW2}$\\ \hline \\

&\multicolumn{6}{c}{Scenario 1: 6 categories, $\beta_1
  = \log(1.5)$, $\beta_2=\log(1.2)$} \\

MC mean & 1.529 & 1.528 & 1.524 & 1.221 & 1.221 & 1.216 \\
MC median & 1.500 & 1.499 & 1.507 & 1.198 & 1.202 & 1.198 \\
MC SD &  0.304 & 0.291 & 0.254 & 0.243 & 0.233 & 0.207 \\
Ave MC SE & 0.298 & 0.285 & 0.249 & 0.239 & 0.228 & 0.199 \\
MC Cov & 0.949 & 0.948 & 0.948 & 0.951 & 0.947 & 0.941 \\
MC MSE ratio & 1.430 & 1.312 & 1.000 & 1.385 & 1.275 & 1.000 \\ 
MC $\pr($reject $H_0)$ & 0.550 & 0.593& 0.700 & 0.156 & 0.165&  0.207 \\*[0.1in]

&\multicolumn{6}{c}{Scenario 2: 6 categories, $\beta_1 = \beta_2 = \log(1.0)$} \\

MC mean & 1.016 & 1.016 & 1.011 & 1.015 & 1.015 & 1.011\\
MC median &  0.997 & 0.999&  0.997 & 0.997 & 0.998 & 0.998\\
MC SD &   0.199 & 0.190 & 0.167 & 0.202 & 0.193 & 0.172 \\
Ave MC SE & 0.195 & 0.187 & 0.163 & 0.198 & 0.189 & 0.165 \\
MC Cov &  0.949 & 0.948 & 0.946 & 0.951 & 0.950 & 0.941  \\
MC MSE ratio &   1.426 & 1.303 & 1.000 & 1.372 & 1.262 & 1.000 \\
MC $\pr($reject $H_0)$ &  0.051 & 0.052 & 0.054 & 0.049 & 0.050 &
                                                                  0.059\\*[0.1in]
  \hline 

\end{tabular}
\end{table}

\label{lastpage}

\end{document}